%% file: main.tex
\documentclass[twocolumn]{aastex63}
\usepackage{float}
\usepackage{graphicx}
\usepackage{amsmath}
\usepackage{natbib}
\usepackage{color}
\usepackage{verbatim}
\usepackage[utf8]{inputenc}
\usepackage{mathtools}
\usepackage{amssymb}
\usepackage{textcomp}
\usepackage{totcount}
\usepackage{enumitem}
\usepackage{braket}
\usepackage{footnote}

\newcommand{\magsqarc}{mag\,arcsec\,$^{-2}$}
\newcommand{\lcdm}{$\Lambda$\textsc{CDM}}

\newtotcounter{citnum} 
\def\oldbibitem{} \let\oldbibitem=\bibitem
\def\bibitem{\stepcounter{citnum}\oldbibitem}

\renewcommand{\thetable}{\arabic{table}}
\shortauthors{Smercina \textit{et al.}}

\makeatletter
\def\blfootnote{\xdef\@thefnmark{}\@footnotetext}
\makeatother

\begin{document}
\title{Relating the Diverse Merger Histories and Satellite Populations of Nearby Galaxies}

\author[0000-0003-2599-7524]{Adam Smercina}
\affiliation{Astronomy Department, University of Washington, Box 351580, Seattle, WA 98195-1580, USA}
\author[0000-0002-5564-9873]{Eric F. Bell}
\affiliation{Department of Astronomy, University of Michigan, Ann Arbor, MI 48109, USA}
\author[0000-0002-8429-4100]{Jenna Samuel}
\affiliation{Department of Astronomy, The University of Texas at Austin, 2515 Speedway, Stop C1400, Austin, TX 78712-1205, USA}
\affiliation{Department of Physics and Astronomy, University of California, Davis, CA 95616, USA}
\author[0000-0001-9269-8167]{Richard D'Souza}
\affiliation{Department of Astronomy, University of Michigan, Ann Arbor, MI 48109, USA}
\affiliation{Vatican Observatory, Specola Vaticana, V-00120, Vatican City State}

\correspondingauthor{Adam Smercina}
\email{asmerci@uw.edu}

\begin{abstract}
We investigate whether the considerable diversity in the satellite populations of nearby Milky Way (MW)-mass galaxies is connected with the diversity in their host's merger histories. Analyzing 8 nearby galaxies with extensive observations of their satellite populations and stellar halos, we characterize each galaxy's merger history using the metric of its most dominant merger, $M_{\rm \star,Dom}$, defined as the greater of either its total accreted stellar mass or most massive current satellite. We find an unexpectedly tight relationship between these galaxies' number of $M_{V}\,{<}\,{-}9$\ satellites within 150\,kpc ($N_{\rm Sat}$) and $M_{\rm \star,Dom}$. This relationship remains even after accounting for differences in galaxy mass. Using the star formation and orbital histories of satellites around the MW and M81, we demonstrate that both likely evolved along the $M_{\rm\star,Dom}$--$N_{\rm Sat}$\ relation during their current dominant mergers with the LMC and M82, respectively. We investigate the presence of this relation in galaxy formation models, including using the FIRE simulations to directly compare to the observations. We find no relation between $M_{\rm\star,Dom}$\ and $N_{\rm Sat}$\ in FIRE, and a universally large scatter in $N_{\rm Sat}$\ with $M_{\rm \star,Dom}$\ across simulations --- in direct contrast with the tightness of the empirical relation. This acute difference in the observed and predicted scaling relation between two fundamental galaxy properties signals that current simulations do not sufficiently reproduce diverse merger histories and their effects on satellite populations. Explaining the emergence of this relation is therefore essential for obtaining a complete understanding of galaxy formation. \\
\end{abstract}

\section{Introduction}
\label{sec:five-intro}

A hallmark of the current $\Lambda$--Cold Dark Matter (\lcdm) paradigm is that the halos of dark matter in which galaxies reside assemble hierarchically \citep[e.g.,][]{white&rees1978}. As a consequence of this hierarchical growth, dark matter and galaxy formation models alike predict that galaxies should accrue extensive populations of lower-mass dwarf `satellite' galaxies, tidally interacting with many of them \citep[e.g.,][]{somerville1999,bullock2001,vandenbosch2002,wechsler2002}. The largest of these accreted galaxies experience strong dynamical friction and can merge with the central galaxy before they tidally disrupt, strongly impacting the properties and evolution of the central galaxy \citep[e.g.,][]{barnes&hernquist1991,hopkins2006}. Modern cosmological simulations, built on a \lcdm\ framework and incorporating these processes, are now able to reproduce many of the most well-constrained large-scale observables, such as the bright end of the galaxy luminosity function \citep[e.g.,][]{bell2003} and the cosmic star formation rate (SFR; \citealt{madau1998}). Perhaps the most important benchmarks of these simulations are the canonical galaxy scaling relations --- for example, the Kennicutt-Schmidt star formation relation \citep[e.g.,][]{kennicutt1998,orr2018}, the Tully-Fisher relation \citep[e.g.,][]{tully&fisher1977,dalcanton1997,bell2001,schaye2015}, and the galaxy star-forming main sequence \citep{noeske2007,schaye2015,jarrett2017} --- which operate across the large range of scales that govern galaxies' properties. Yet, despite this impressive progress in explaining the formation and evolution of relatively massive galaxies, the regime of small-scale galaxy formation has remained problematic for simulations. 

This difficulty in modeling small-scale galaxy formation is due, in large part, to the sensitivity of low-mass galaxy properties to stellar feedback and other baryonic processes. Consequently, this results in a magnification of small differences in physical prescriptions on predicted galaxy properties in the low-mass regime \citep[e.g.,][]{bullock&boylan-kolchin2017}. As processes such as stellar feedback can operate on both extremely small spatial and short time scales, simultaneously achieving the required spatial and temporal resolution for the complex associated physics is extremely difficult \citep[e.g.,][]{hopkins2014}. Thus, modeling galaxy formation at the smallest scales, and therefore investigating the model-based links between small- and large-scale galaxy formation, has proved challenging to date.

The persistent difficulty of modeling small-scale galaxy formation has motivated two decades of careful theoretical and observational study of dwarf galaxies. Difficult to detect observationally owing to their intrinsic faintness, dwarf galaxies are both the lowest-mass and most dark matter-dominated galaxies in the Universe \citep[e.g., see the recent review by][]{wechsler&tinker2018}. Consequently, their properties are critical benchmarks for our models of galaxy formation. Dwarf galaxies are the observational bedrock of our understanding of small-scale cosmology \citep[e.g., see the recent review by][]{bullock&boylan-kolchin2017}. 

While they are predicted to also exist in isolation, until recently nearly all of the low-mass dwarf galaxies readily accessible for observational study were `satellites' of Milky Way (MW)-mass galaxies \citep[e.g., see][for an overview of the Local Group satellite populations]{mcconnachie2012} --- existing within an environment which is dominated by a much larger central galaxy. This is an important consideration, as the large-scale environments of galaxies like the MW are not static, `closed' boxes --- they are complex ecosystems that frequently experience substantial mergers with other systems throughout their lives.

Are galaxies like the MW expected to have experienced similar ensemble merger histories over the course of their lives? Quite the opposite! Theoretical modeling has long predicted that MW-mass galaxies should have a wide range in their merger histories, and particularly in the largest mergers that they have experienced \citep[e.g.,][]{purcell2010,deason2015}, with many galaxies experiencing mergers with progenitors more massive than the Large Magellanic Cloud (LMC; \citealt{stewart2008}). Study of their stellar halos has largely confirmed this predicted diversity by observing a wide range of stellar halo masses and metallicities around galaxies with masses comparable to the MW, with correlations between them showing that stars accreted from the largest past merger dominate the properties of the halo stellar populations \citep{harmsen2017,bell2017,dsouzabell2018a,smercina2020}. Meanwhile, recent evidence suggests that the `classical' satellite populations of nearby MW-analogs are equally diverse \citep{geha2017,smercina2018,bennet2019,carlsten2021}.

Recent evidence from the Local Group (LG) suggests that the diversity in both of these fundamental properties of galactic systems may be related. Studies of the MW's satellites using \textit{Gaia} data and detailed orbital modeling postulate that a number of its satellites may have been brought in during the infall of the Large Magellanic Cloud system \citep[e.g.,][]{gaiacollaboration-brown2018,kallivayalil2018,jahn2019,pardy2020,erkalbelokurov2020,patel2020} --- possibly including some of the MW's `classical' satellites, such as Carina and Fornax. Additionally, deep resolved star formation histories of the M31's dwarf satellites with the \textit{Hubble Space Telescope (HST)} have recently revealed that nearly 50\% share a remarkably similar timescale for the `shutdown' of their star formation, $\sim$3--6\,Gyr ago \citep{weisz2019}. This is similar to the predicted first infall time of the massive progenitor galaxy whose merger with M31 likely formed M31's massive and structured stellar halo \citep[M32p;][]{hammer2018,dsouzabell2018b}. This suggests that the merger histories of the MW and M31 \textit{may need to be accounted for} in order to explain the properties of their present-day satellite populations.

This represents an important new lens through which to view and test predictions for MW-like systems from galaxy formation models. Is the infall of satellites during mergers with galaxies such as the LMC or M32p surprising? In a dark matter only (DMO) view, the overall number of surviving subhalos within a central halo will primarily be a function of the mass of the central halo \citep[e.g.,][]{kravtsov2004}, similar to what has long been seen in the idealized case of rich galaxy clusters \citep[e.g.,][]{bahcall-cen1993}. However, due to the self-similarity of CDM, dwarf galaxies are predicted to host their own populations of satellites prior to group infall \citep[e.g.,][]{deason2015b,dooley2017b,dooley2017} and there has been direct confirmation of these `satellites-of-satellites' in both the Local and nearby groups \citep[e.g.,][]{deason2014,deason2015,smercina2017}. Therefore, these large satellites are entirely expected to bring in their own populations of low-mass satellites. Indeed, it is expected that the distribution of satellite infall times in MW-like halos should cluster around massive accretion events \citep[e.g.,][]{deason2015b,dsouzabell2021}. Yet, because of the strong halo mass dependence on the number of preexisting low-mass satellites, and the fact that \textit{all} galaxies experience accretions, the default DMO prediction is that the satellite populations of MW-mass galaxies should not, overall, correlate strongly with their late-time accretion history \citep{bose2020}.

However, the baryonic physics involving infalling satellites' interactions with circumgalactic gas, tidal disruption by the central galaxy, and internal stellar feedback complicates matters significantly \citep[e.g.,][]{wetzel2015}. Understanding the importance of merger history in predicting the properties of present-day satellite populations requires both extremely high resolution and a sample of MW-mass systems with diverse merger histories --- a set of constraints that are only now being achieved in modern simulations \citep[e.g.,][]{simpson2018,garrison-kimmel2019,monachesi2019,samuel2020,engler2021}. Given this theoretical progress, it is therefore important to observationally explore how merger history relates to satellite populations, and compare it to our theoretical predictions. 

\input{obstbl}

In this paper, we leverage the recent wealth of observational insight into the merger histories and satellite populations of MW-mass galaxies to empirically investigate a connection between them in nearby systems, with the goal of providing an observational benchmark for galaxy formation models. The paper is structured as follows: in \S\,\ref{sec:bg} we first outline the existing observational data for both satellite populations (\S\,\ref{sec:sat-bg}) and stellar halo measurements (\S\,\ref{sec:sh-bg}) around MW-like galaxies, and the intuition that has been gained from each. In \S\,\ref{sec:satacc} we present the results of our initial comparison between the most dominant merger and the number of classical satellites in eight nearby systems, which exhibit a clear relationship. In \S\,\ref{sec:significance} we quantify the form and significance of this relationship (\S\,\ref{sec:model-define}), accounting for an intrinsic scaling between satellite populations and galaxy mass (\S\,\ref{sec:stmass}). In \S\,\ref{sec:built}, we demonstrate the evolution of the MW's and M81's satellite populations along this relation. In \S\,\ref{sec:model-comparison} we compare the observed scaling relation to theoretical predictions from the FIRE simulations, followed by an in-depth discussion of the implications for galaxy formation and possible origins of the relation in \S\,\ref{sec:implications}. We summarize our conclusions in \S\,\ref{sec:conclusions}. 

\section{The Current Census of Milky Way-like Systems}
\label{sec:bg}

Addressing a possible connection between the satellite populations and merger histories of MW-mass galaxies requires measurement of two notoriously difficult-to-measure properties: (1) global satellite populations, and (2) global stellar halo properties. Efforts to study both of these properties for MW-mass systems have remained a substantial focus of the field over the last two decades. However, they have been historically kept in separate `intellectual boxes'. In this paper, we combine these insights to ask the question: are galaxies' merger histories and satellite populations connected? As the vast majority of existing data has been obtained in MW-mass systems, this is where we concentrate our efforts. In this section we summarize the current state of the field regarding the satellites (\S\,\ref{sec:sat-bg}) and stellar halos (\S\,\ref{sec:sh-bg}) of nearby MW-mass galaxies.

\subsection{The Diverse Satellite Population of MW-mass Galaxies}
\label{sec:sat-bg}

Tensions in the number and properties of observed MW satellites, relative to model predictions --- e.g., the `Missing Satellites' and `Too Big to Fail' problems \citep{klypin1999,moore1999,boylan-kolchin2011} --- constitute some of the largest hurdles for the \lcdm\ paradigm \citep[e.g.,][]{bullock&boylan-kolchin2017}. Solutions to these problems --- often focusing on the impact of baryonic processes such as reionization \citep[e.g.,][]{bullock2001} and stellar feedback \citep[e.g.,][]{brooks2013} --- have used the satellite population of the MW as a benchmark. Just this one system has been used to direct the scope of some of the most important problems in galaxy formation. 

Motivated by this lack of context, the field's focus has shifted to surveying the satellite populations of nearby MW-analogs (i.e.\ central galaxies in the Local Volume with stellar mass $M_{\star}\,{\sim}$\,3--20${\times}10^{10}\,M_{\odot}$). Through a combination of wide-field integrated light searches, targeted resolved star studies, and spectroscopic surveys these efforts have greatly enhanced our understanding of satellite galaxy populations. In all, eight MW-mass systems have now been surveyed to the depth of the MW's `classical' satellite population ($M_{V}\,{\lesssim}\,{-}9$): the MW and M31 \citep[compiled by][]{mcconnachie2012}, M81 \citep[compiled by][]{karachentsevkudrya2014}, M101 \citep{danieli2017,bennet2019,carlsten2019}, Centaurus A \citep{crnojevic2019}, M94 \citep{smercina2018}, M83 \citep[e.g.,][]{muller2015,muller2017,carrillo2017}, and M104 \citep{carlsten2021}. 

\begin{figure*}[t]
\leavevmode
\centering

\begin{minipage}{0.43\textwidth}
    \includegraphics[width={\linewidth}]{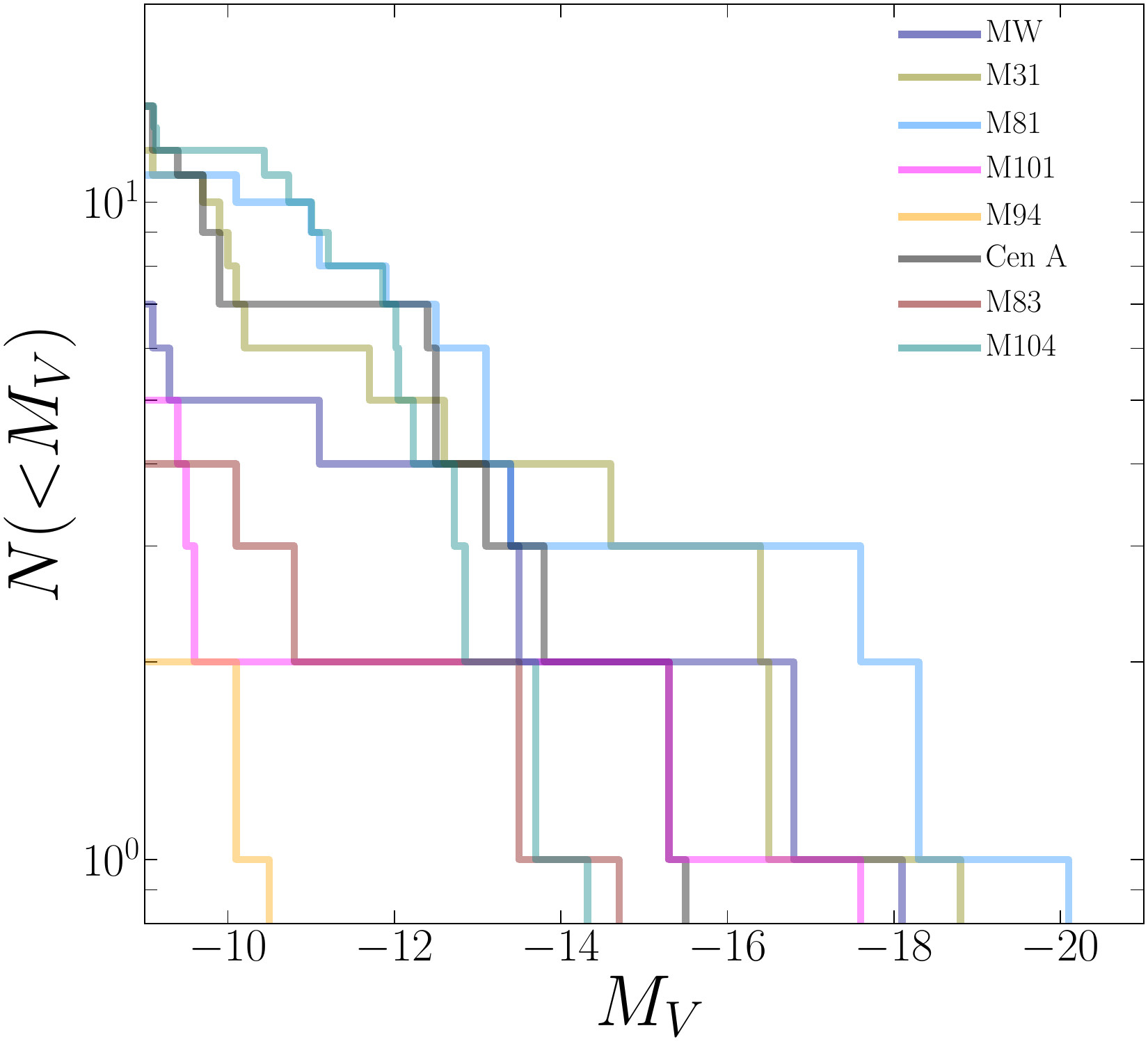}
\end{minipage}
\begin{minipage}{0.55\textwidth}
    \includegraphics[width={\linewidth}]{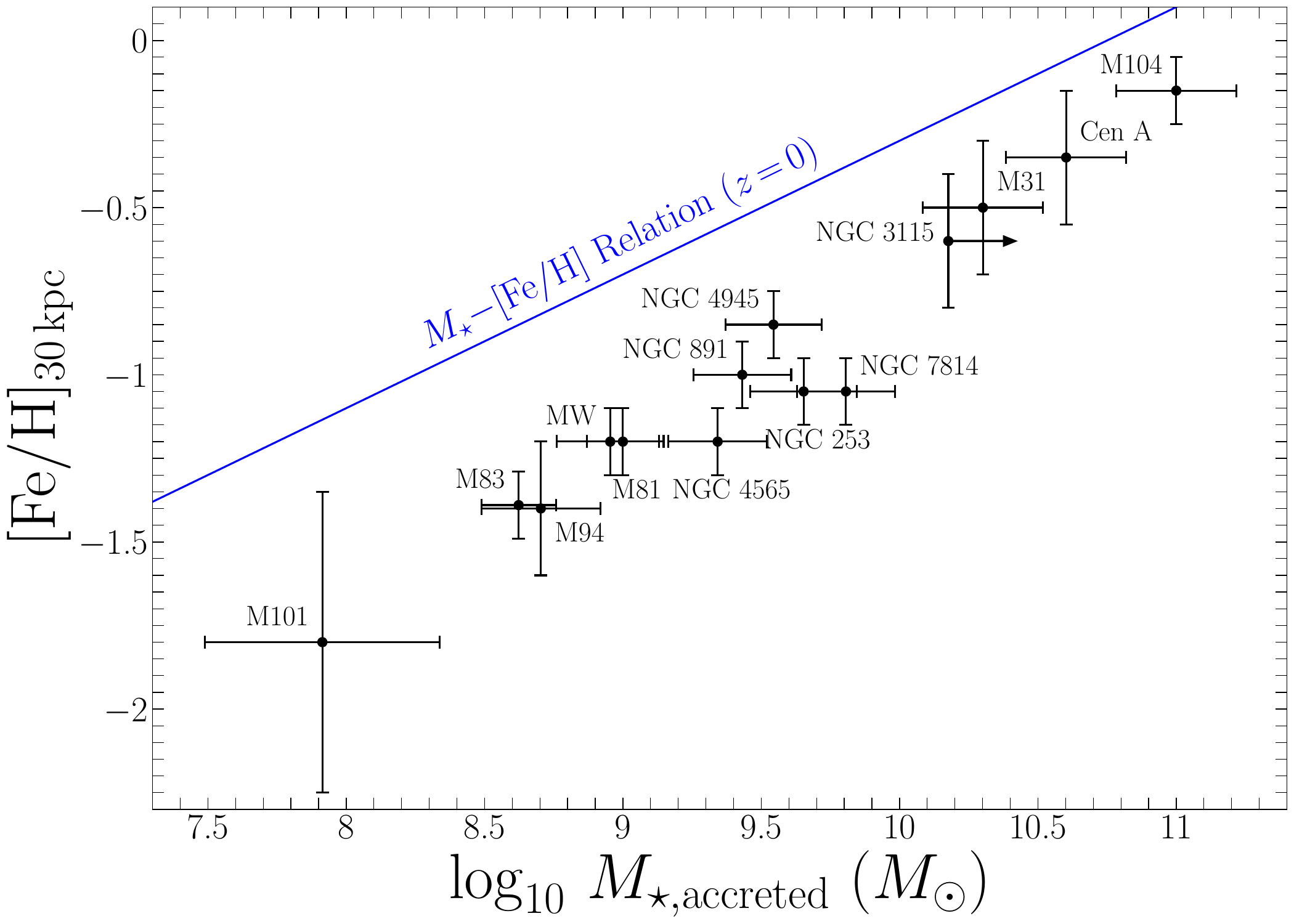}
\end{minipage}

\caption{\textbf{\uline{Left}:} Cumulative $V$-band satellite luminosity functions, within a projected 150\,kpc galactic radius, for the eight MW-mass systems which are complete to $M_V\,{\lesssim}\,{-}9$\ (i.e. `classical' satellites). Adapted and from \cite{smercina2018} and updated with recent work (see Table \ref{tab:obs} for references). The luminosity function for M104 is an average of the upper- and lower-limit cases from \cite{carlsten2021}. Highlighted particularly well by the sparse satellite population of the `lonely giant' M94, these eight systems showcase a broad diversity in the satellite populations of galaxies at the MW-mass scale. \textbf{\uline{Right}:} The stellar halo mass–metallicity relation for the Milky Way and 13 other nearby MW-mass galaxies, including the eight shown in the left panel. Total accreted mass ($M_{\rm \star,accreted}$), estimated following \cite{harmsen2017} (see \S\,\ref{sec:sh-bg}), is plotted against metallicity measured at 30\,kpc ([Fe/H]$_{\rm 30\,kpc}$). See Table \ref{tab:obs} for references from which the data were compiled. An approximation of the $z\,{=}\,0$\ stellar mass--metallicity relation \citep{gallazzi2005,kirby2013} is shown in blue for reference. The broad range of stellar halo properties displayed here --- three orders of magnitude in mass and nearly two in metallicity --- primarily indicates a broad range in the most dominant mergers these galaxies have experienced.}
\label{fig:sat-sh_census}
\end{figure*}

With this newfound access to a true sample of satellite populations in MW-like galactic systems has come the realization that these populations are incredibly diverse \citep{geha2017,smercina2018,bennet2019,carlsten2021}. This diversity, particularly the discovery of the sparse satellite population around the `lonely giant' M94, is a particularly powerful constraint on small-scale galaxy formation, including the low-mass end of the stellar mass--halo mass (SMHM) relation \citep{smercina2018,carlsten2021}. Figure \ref{fig:sat-sh_census} (left) shows the satellite $V$-band luminosity functions (LFs) within 150\,kpc projected galactic radius for the eight MW-mass galaxies that have been studied down to $M_V\,{\lesssim}\,{-}9$. We note that most of the satellites in these eight systems now have existing distance estimates (most from \textit{HST} color--magnitude diagrams) and therefore their line-of-sight (LOS) distances can largely be constrained to within $\pm$1\,Mpc of the central galaxies, except in the cases of M101 and M104 (described below). Given this, we apply a 1\,Mpc constraint, in addition to the 150\,kpc projected radius, following \cite{smercina2018}. This is important for the MW and M31 where we take the \textit{projected} satellite LFs, calculated from random LOSs within each. Another notable consequence is that NGC 5253 is excluded as an M83 satellite, due to its likely foreground status.\footnote{Distances: 4.66\,Mpc for M83 vs.\ 3.55\,Mpc for NGC 5253 \citep[Extragalactic Distance Database;][]{tully2013}}.  

There has been significant back-and-forth discussion in the literature about M101's satellite population, regarding which satellite candidates are truly associated with M101 vs.\ in the background, as well as their integrated brightnesses \citep{muller2017b,danieli2017,bennet2019,carlsten2021}. \cite{bennet2019} determined that a number of the satellites considered by \cite{danieli2017} were actually background objects, while \cite{carlsten2021} added three satellites (eight vs.\ five) compared to \cite{bennet2019}, due to a claimed discrepancy in their $V$-band luminosities. For this paper we adopt \cite{bennet2019} as M101's fiducial LF, given its similarity to the other surveys around nearby systems, but incorporate the \cite{carlsten2021} LF into our uncertainties. The \cite{carlsten2021} LF for M104 is also relatively uncertain, as surface brightness fluctuation (SBF) distance estimates are inconclusive for a number of its satellites. Given the richness of its satellite population, we consider it to be `measured', despite a range of faint-end LFs. The M104 LF shown in Figure \ref{fig:sat-sh_census} is an average of the lower- and upper-bound LFs in each magnitude bin.  

\subsection{Inferring a Galaxy's Most Dominant Merger from its Stellar Halo Properties}
\label{sec:sh-bg}

With near-equal vigor to the efforts to study the satellite populations of MW-analogs, there has been a significant push to extract information about these galaxies' merger histories. Motivated by insight from galaxy formation models, suggesting that MW-mass galaxies likely experience diverse merger histories \citep[e.g.,][]{bullock&johnston2005,purcell2007}, such efforts have sought to use these galaxies' stellar halos as probes of past merger events. 

Comparisons between the observed properties of MW-analogs' stellar halos and galaxy formation models suggests that these halos are primarily composed of the disrupted remnants of accreted satellites \citep[e.g.,][]{harmsen2017,dsouzabell2018a,monachesi2019}. Further, models predict that the measurable properties of these accreted stellar populations are often dominated by the most massive merger the central has experienced \citep[e.g.,][]{deason2015b,dsouzabell2018a,monachesi2019}. Recent detailed studies of the halos of M31 \citep{dsouzabell2018b} and M81 \citep{smercina2020} support this picture. Since the stellar mass present in the stellar halo is primarily accreted from the largest merger partner, the measured accreted mass is approximately equivalent to the stellar mass of the dominant progenitor galaxy. 

Observed samples of galaxies for which stellar halo properties have been well-measured now exist \citep[e.g.,][]{merritt2016,monachesi2016a,harmsen2017}. A primary realization from these surveys is that stellar halos of MW-mass galaxies are very diverse and form a crude-but-powerful stellar halo mass--metallicity relation \citep{harmsen2017}. This relation naturally emerges if the accreted material is dominated by the underlying stellar mass--metallicity relation present in the constituent satellites, albeit with a different zero-point.\footnote{At a given stellar mass, the stellar halo metallicity is lower than the present-day galaxy stellar metallicities owing to both metallicity evolution in galaxies and dilution of metallicities by (sub-dominant) contributions to stellar halos from low-mass more metal-poor galaxies \citep{harmsen2017,dsouzabell2018a,monachesi2019}.} The existence of this stellar halo mass--metallicity relation, across a broad range of accreted mass, is therefore further observational confirmation that the stars in a stellar halo are dominated by the most massive accreted satellite. Figure \ref{fig:sat-sh_census} shows the inferred total accreted stellar mass plotted against inferred photometric metallicity measured at 30\,kpc along the minor axis for 14 galaxies in the Local Volume for which both properties have been measured (or have robust limits). Total accreted stellar mass has been estimated for each galaxy, following \cite{harmsen2017}, by integrating the star count-scaled projected 2-D density profile in the range of 10--40\,kpc and multiplying by a factor of three --- obtained from comparisons to models \citep{bullock&johnston2005,bell2017}. The metallicity at 30\,kpc is inferred from the minor axis average metallicity profile of the resolved stars, obtained via comparison to stellar population models. 

Publications presenting the accreted mass estimates for two of the galaxies shown here are still in preparation, so we summarize the results here. M83's stellar halo has been measured using observations from the GHOSTS survey \citep{radburn-smith2011}. M83 has an extensive envelope of metal-poor RGB stars and a well-known stellar stream \citep{malin&hadley1997,barnes2014}, both of which have been analyzed \citep[following][]{harmsen2017} by \cite{cosby2022} to obtain an estimated accreted mass of $M_{\star}\,{=}\,4.2{\times}10^8\ M_{\odot}$, with an average metallicity of [Fe/H]\,$=\,{-}1.40$\ at 30\,kpc. Observations of M94's stellar halo exist from both GHOSTS and observations with the Subaru Hyper Suprime-Cam \citep[HSC;][]{smercina2018}, which are of very similar quality to those in M81's stellar halo \citep{smercina2020}. Using comparable analysis to the study of M81's global halo (including estimating photometric completeness from artificial star tests), \cite{gozman2022} use RGB star counts to estimate a total accreted mass for M94 of $M_{\star}\,{=}\,5.05{\times}10^8\,M_{\odot}$, with an average metallicity of [Fe/H]\,$=\,{-}1.4$\ at 30\,kpc.

As can be seen in Figure \ref{fig:sat-sh_census}, nearby MW-mass galaxies display an enormous range of accreted properties, ranging from M101's accreted mass of ${<}10^8\,M_{\odot}$\ to M104's ${\sim}10^{11}\,M_{\odot}$. Though exhibiting a steeper overall slope, likely indicating relevant contributions from several accretion events at the low-mass end and more recent merger times predicted for the largest mergers \citep{dsouzabell2018b}, these galaxies' stellar halo properties closely track the stellar mass--metallicity relation. The implied dominant mergers that these galaxies have experienced range from the stellar mass of the Fornax dwarf galaxy to the stellar mass of M31. Given the hierarchical nature of galaxy formation, it is important to understand how these dramatic differences in merger history may impact system-wide galaxy properties, and whether they could help to explain the observed diversity in these galaxies' satellite populations. 

\begin{figure*}[t]
\leavevmode
\centering
\includegraphics[width={0.8\linewidth}]{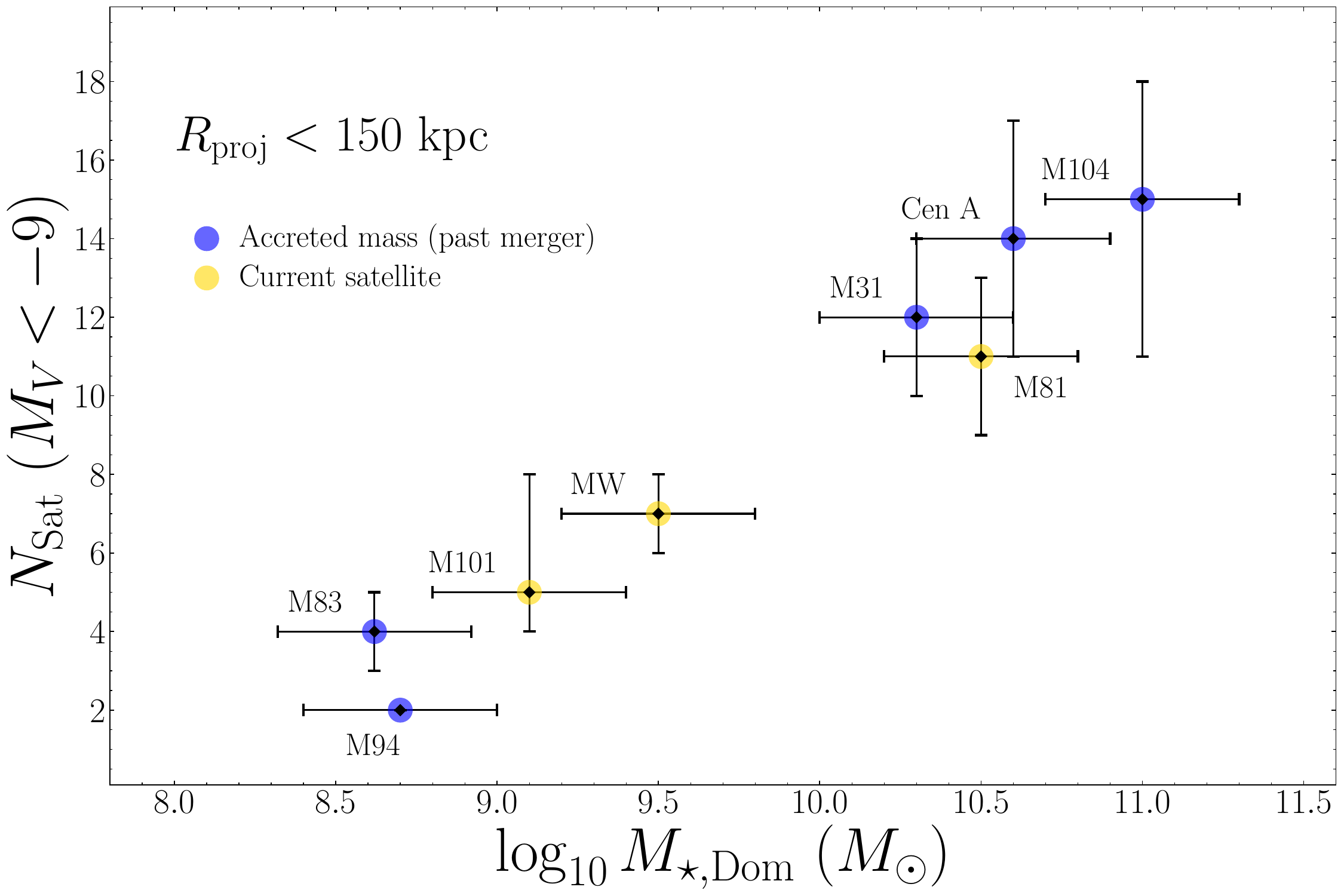}
\caption{Total number of satellites with $M_V\,{<}\,{-}9$, within 150\,kpc projected radius, around each of eight nearby MW-mass galaxies, plotted against the mass of the most dominant merger they have experienced (see \S\,\ref{sec:model-comparison}). Galaxies are color-coded according to whether $M_{\rm \star,Dom}$\ reflects the accreted material from a past merger (blue), or the mass of an existing satellite (gold). See Table \ref{tab:obs} for the data represented here. A clear linear relationship is visible.
}
\label{fig:sat-accretion}
\end{figure*}

\section{Comparing Nearby Galaxies' Most Dominant Mergers and Satellite Populations}
\label{sec:satacc}

Through these efforts, a sample of eight systems  --- the MW, M31, M81, M83, M94, M101, M104, and Cen A --- now exists for which both complete satellite LFs and accreted masses have been measured. These datasets present a brand new and unprecedented opportunity to explore a connection between these two fundamental properties of nearby well-studied galactic systems.

First, we will define several quantities which will be used in this analysis. As discussed in \S\,\ref{sec:sh-bg}, the most dominant merger a galaxy has experienced is well estimated by its total accreted stellar mass. However, in a number of systems the most dominant merger is still in progress, in the form of a nearby massive satellite. In this case the galaxy `ecosystems' (i.e. dark matter halos) have already begun to merge, but the galaxies have yet to coalesce. The most familiar example of this is the MW itself: as described in \S\,\ref{sec:five-intro} a number of MW satellites may have originally been satellites of the (not-yet-disrupted) Large and Small Magellanic Clouds (LMC and SMC). Additionally, recent evidence suggests that the LMC's massive halo has already had a substantial global impact on the MW's halo \citep{conroy2021}, suggesting that they have indeed begun to merge on a large scale. 

We contend that the most massive merged ecosystem is likely the important property to consider for satellite populations, and while the accreted mass is useful as a proxy for identifying past such mergers, the most massive existing satellite is an equally powerful proxy of current ongoing mergers, even if the galaxies have not yet interacted significantly. For example, \citet{smercina2020} showed that in systems such as M81, which is in the early stages of a close merger with its massive satellite M82, the merger is not advanced enough to have redistributed the accreted material into the stellar halo (the estimate of the current accreted mass in M81's stellar halo is $\sim$1/20 of its estimated total future accreted mass). Yet given M81's modest accreted mass, M82 undoubtedly represents the most massive \textit{ecosystem} merger that M81 has experienced. In light of this, we instead adopt a revised metric for the mass of the most dominant \textit{ecosystem} merger, $M_{\rm \star,Dom}$, where the proxy is the larger of the total accreted mass or the mass of the most massive satellite. We adopt uniform conservative 0.3\,dex uncertainties on $M_{\rm \star,Dom}$\ \citep[following][]{harmsen2017,bell2017,dsouzabell2018a}, reflecting the large uncertainties on both the inferred stellar mass of nearby galaxies and model-dependent accreted mass estimates.

For satellites, we only have access to the `full' classical satellite populations (i.e.\ $M_{V}\,{<}\,{-}9$\ satellites out to the virial radius) of the MW and M31. We therefore follow recent studies \citep[e.g.,][]{smercina2018,bennet2019,carlsten2021} and restrict our attention to satellites within a projected galactic radius of 150\,kpc, down to an absolute $V$-band magnitude of $M_{V}\,{<}\,{-}9$\ (see Figure \ref{fig:sat-sh_census}). Though there is diversity in both survey area and depth, all eight systems are considered complete within these radius and brightness criteria. 

Figure \ref{fig:sat-accretion} shows $M_{\rm \star,Dom}$\ plotted against $N_{\rm Sat}$\ for the eight available MW-mass systems, color-coded by whether $M_{\rm \star,Dom}$\ represents a past merger (i.e. accreted mass; blue) or an existing satellite (yellow).\footnote{See \textsc{appendix} \ref{sec:m51-ngc4565} for limits on M51 and NGC 4565.} A clear and surprisingly tight relationship is visible, spanning the entire three orders of magnitude in dominant merger mass --- systems that have experienced larger mergers host more satellites. 

\section{Significance of the Relationship}
\label{sec:significance}

In this section, we will quantify the significance of the observed $N_{\rm Sat}$--$M_{\rm \star,Dom}$\ relation. In \S\,\ref{sec:model-define} we define our adopted model to describe the relation, followed by the results of an MCMC analysis. In \S\,\ref{sec:stmass} we consider and account for a possible relationship between galaxy stellar mass and number of satellites, showing that the relationship remains in the residuals of an imposed $M_{\rm \star,Gal}$--$N_{\rm Sat}$\ relation. 

\begin{figure*}[t]
\leavevmode
\centering

\begin{minipage}{0.52\linewidth}
    \includegraphics[width={\linewidth}]{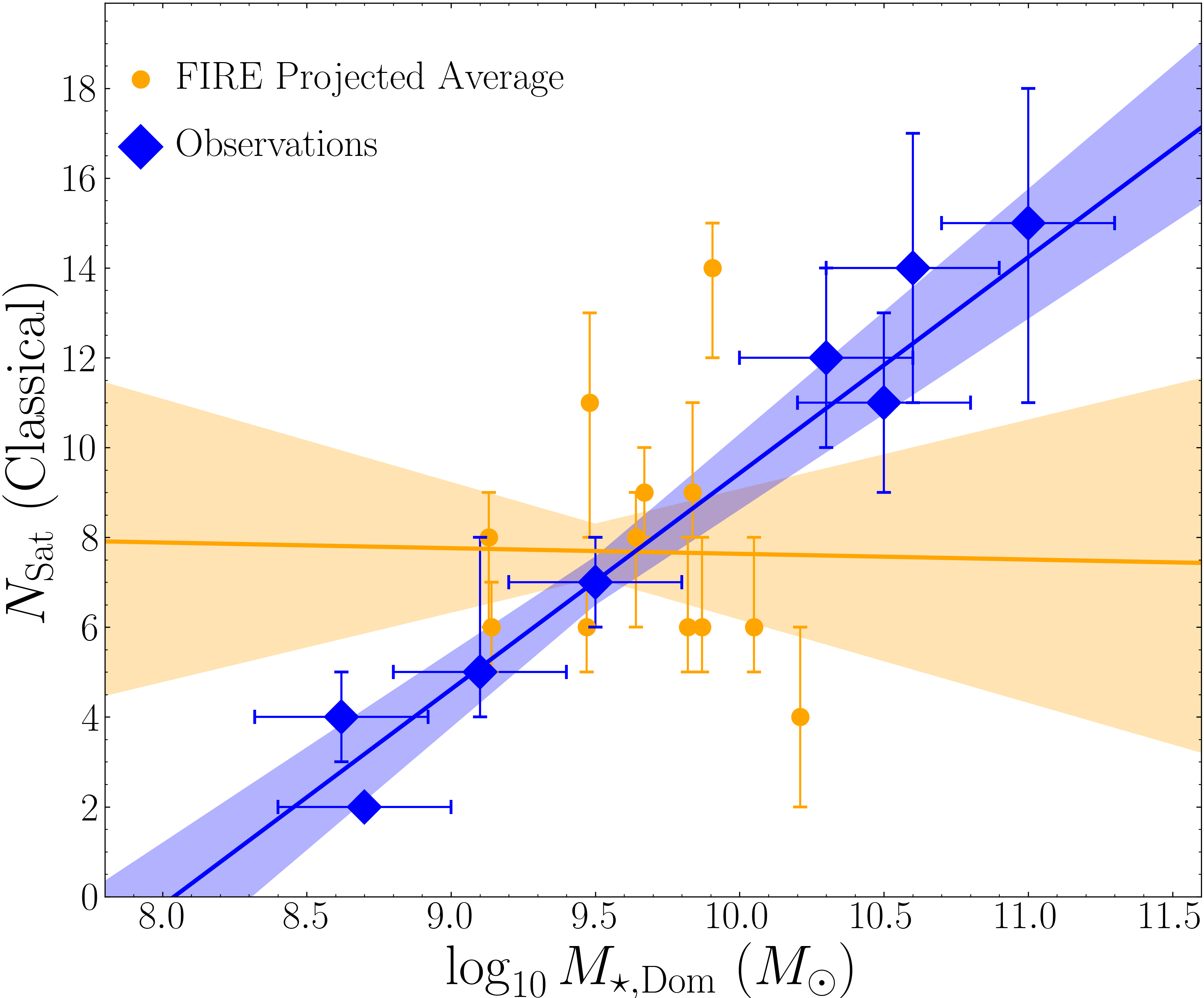}
\end{minipage}
\begin{minipage}{0.45\linewidth}
    \includegraphics[width={\linewidth}]{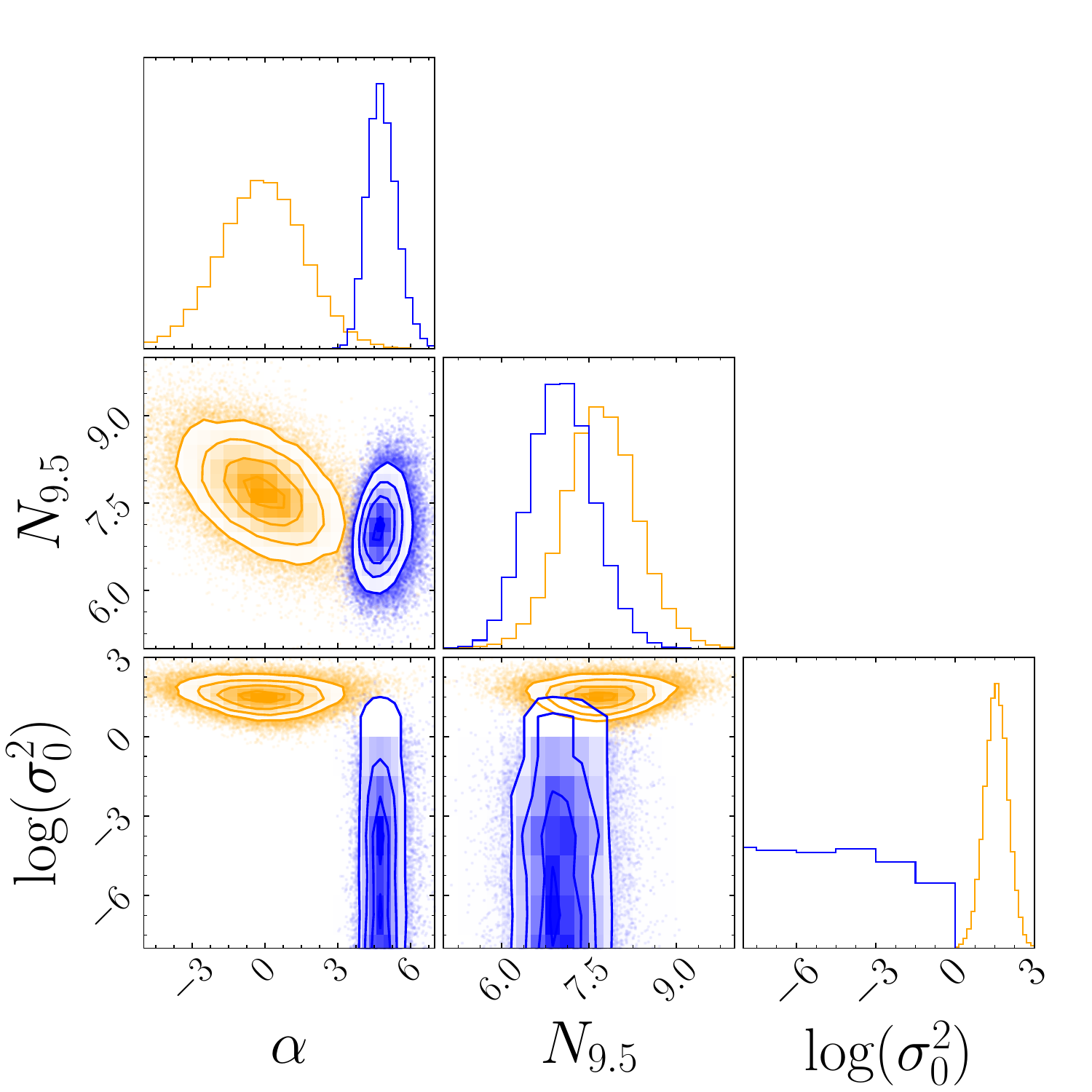}
\end{minipage}

\caption{\uline{Left}: A comparison of the linear model parameters that best describe the eight observed systems (blue) and systems from FIRE (orange), with satellite populations estimated within 150\,kpc projected radius. We show the most likely models from our \texttt{emcee} analysis for both FIRE and the observed sample, as well as the 16--84\% confidence regions for each. \uline{Right}: A \texttt{corner} plot showing the posterior distributions from our \texttt{emcee} analysis for all model parameters, given both the FIRE (orange) and observed (blue) data. Note that we present the \textit{natural logarithm} of $\sigma_0^2$, and that we cut off the posterior distribution for $\log(\sigma_0^2)$\ at $-8$, as it just tracks the flat prior (${-}20\,{<}\,\log(\sigma_0^2)\,{<}\,0$) for $\log(\sigma_0^2) < {-}3$.}
\label{fig:FIRE-compare}
\end{figure*}

\subsection{Defining the Model}
\label{sec:model-define}

For the purposes of simplicity, we will parameterize the relationship between $M_{\rm \star,Dom}$\ and $N_{\rm Sat}$\ as a simple linear model, given by
\begin{equation}
    N(\alpha,N_{9.5};M_{\rm \star,Dom}) = \alpha (\log_{10}{M_{\rm \star,Dom}} - 9.5) + N_{9.5},
    \label{eq:1}
\end{equation}
where we have normalized the model to the MW's value of $\log_{10}{M_{\rm \star,Dom}} = 9.5$\ (see Table \ref{tab:obs}). To estimate the probability of the data given our assumed model we adopt the appropriate likelihood function for a linear model with intrinsic scatter \citep[following][]{hoggbovylang2010}:
\begin{equation}
    \ln \mathcal{L} = -\frac{1}{2} \sum_{i=1}^n \frac{[N_{i,{\rm Sat}} - N_i(\alpha,N_{9.5};M_{\rm \star,Dom})]^2}{\sigma_i^2} + \ln(\sigma_i^2), 
    \label{eq:2}
\end{equation}
where the variance $\sigma_i^2$\ is expressed as 
\begin{equation}
    \sigma_i^2 = (\alpha\,\sigma_{\log_{10}M_{{\rm \star,Dom},i}})^2 + \sigma_{N_{{\rm Sat},i}}^2 + \sigma_0^2.
    \label{eq:3}
\end{equation}
$\sigma_i^2$, defined for each system $i$, takes into account uncertainties on both $M_{\rm \star,Dom}$\ and $N_{\rm Sat}$, $\sigma_{\log_{10} M_{{\rm \star,Dom},i}}$\ and $\sigma_{N_{{\rm Sat},i}}$\ respectively, as well as an intrinsic scatter term which is assumed to be the same for all systems, $\sigma_0^2$. 

We compute the posterior probability distributions for each model parameter given the data using the Python-based Markov chain Monte Carlo (MCMC) sampler \texttt{emcee} \citep{foreman-mackey2013}. Figure \ref{fig:FIRE-compare} (left panel) shows data plotted against the likeliest model in blue, as well as the 16--84\% confidence regions. We also show the marginalized single and joint posterior distributions for the model parameters given the data (right panel, again in blue). For comparison, Figure \ref{fig:FIRE-compare} also shows results from the FIRE simulation suite; we defer description and discussion of this comparison  to \S\,\ref{sec:model-comparison}.

The eight galaxies are best described by a tight relation, with intrinsic scatter well within the uncertainties (in fact, consistent with zero scatter in our best-fit model). The model with the highest likelihood of describing the observations is:
\begin{equation}
    N_{\rm Sat} = 4.81_{-0.56}^{+0.64}(\log_{10}M_{\rm \star,Dom}-9.5) + 7.02_{-0.53}^{+0.55}.
\end{equation}

\subsection{Stellar Mass Considerations}
\label{sec:stmass}

\begin{figure*}
\leavevmode
\centering

\hspace{-15pt}
\begin{minipage}{0.29\linewidth}
    \includegraphics[width={\linewidth}]{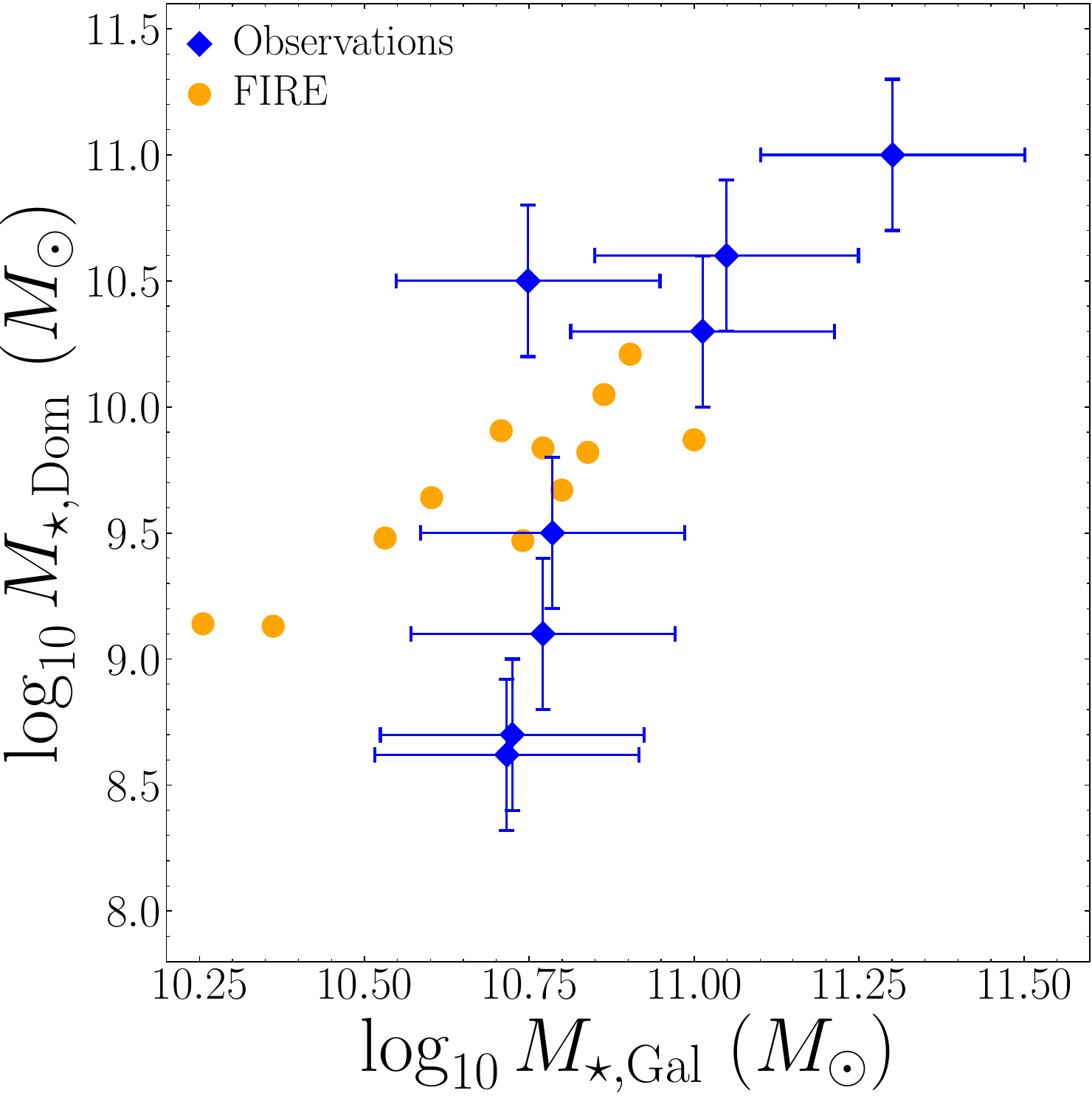}
\end{minipage}
\begin{minipage}{0.29\linewidth}
    \includegraphics[width={\linewidth}]{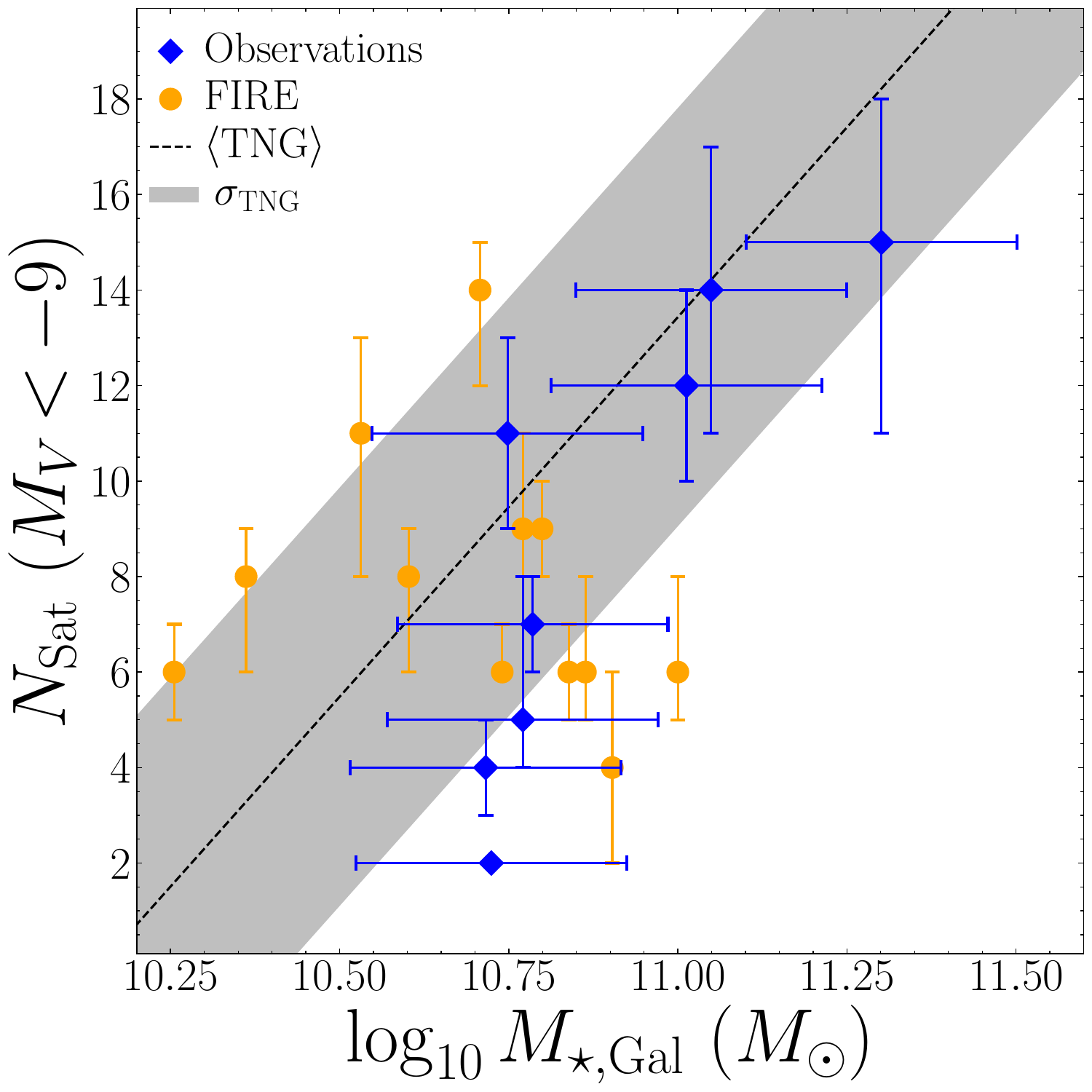}
\end{minipage}
\hspace{-6pt}
\begin{minipage}{0.37\linewidth}
    \includegraphics[width={\linewidth}]{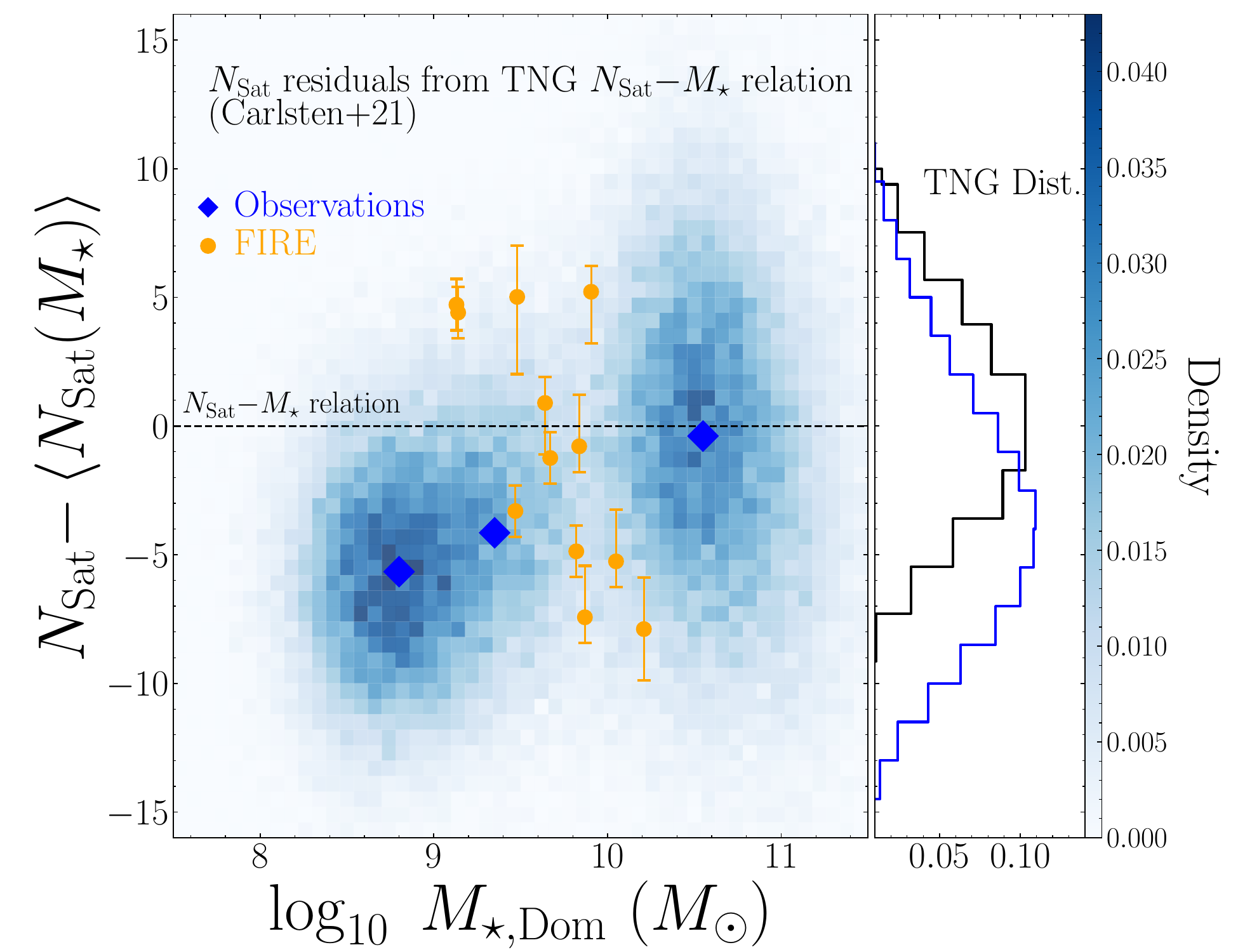}
\end{minipage}

\caption{This figure shows the persistence of the $M_{\rm \star,Dom}$--$N_{\rm Sat}$\ correlation even after accounting for intrinsic correlations with galaxy stellar mass. \uline{Left}: $M_{\rm \star,Gal}$\ vs.\ $M_{\rm \star,Dom}$\ for the observed (blue) and FIRE (orange) samples. The eight galaxies do not show a strong correlation. Surprisingly, the FIRE systems show a stronger positive correlation. \uline{Center}: $M_{\rm \star,Gal}$\ vs.\ $N_{\rm Sat}$\ for the observed and FIRE samples. Also shown as a dashed line is the average relation found by \cite{carlsten2021}, as well as the RMS scatter they measure (shaded gray). The eight observed galaxies are modestly correlated, though this is driven by the two most massive galaxies. The FIRE systems do not exhibit any correlation and largely fall outside of the RMS range from TNG. \uline{Right}: The results of our bootstrap Monte Carlo tests (see \S\,\ref{sec:stmass}), binned by 0.08 in $\log_{10}M_{\rm \star,Dom}$\ and 0.6 in $N_{\rm Sat}{-}\left<N_{\rm Sat}\right>$, is shown in blue. We show the average $N_{\rm Sat}{-}\left<N_{\rm Sat}\right>$\ for each of the three visible clusters in $M_{\rm \star,Dom}$\ as solid blue diamonds. We also show the FIRE sample as orange filled circles. We find that the average $N_{\rm Sat}$\ residual in each $M_{\rm \star,Dom}$\ bin is correlated with $M_{\rm \star,Dom}$\ for the observed sample, though see no correlation for the FIRE sample. To the right we show the distributions of $N_{\rm Sat}{-}\left<N_{\rm Sat}\right>$\ from the abundance-matched Illustris-TNG analysis (black; \citealt{carlsten2021}) and the randomly-sampled observations (blue). The distributions have a comparable shape, however the observations are systematically shifted to negative $N_{\rm Sat}{-}\left<N_{\rm Sat}\right>$\ values. 
}
\label{fig:sat-residuals}
\end{figure*}

An important consideration in explaining the emergence of this relation is the range of stellar masses of the observed galaxy sample. New evidence suggests that the stellar mass of central galaxies may correlate with how many satellites they host \citep{carlsten2021}, reflecting the non-zero slope of the SMHM relation at the `MW mass' --- i.e. galaxies with higher stellar masses should live, on average, in more massive DM halos and thus are predicted in hierarchical assembly to host more sub-halos. Of the eight observed galaxies, only the MW and M31 have reliable estimates of their halo masses, inferred from the motions of individual stars \citep[e.g.,][and many others]{cautun2020,deason2021} or their full satellite populations \citep[e.g.,][]{watkins2010,patel2018}, thus we cannot directly investigate the role of halo mass in setting $N_{\rm Sat}$. We can, however, follow precedents set in the literature \citep[e.g.,][]{carlsten2021} and use galaxy stellar mass as a (scattered) proxy for halo mass to assess the possible intrinsic impact of galaxy mass on satellite number.

In Figure \ref{fig:sat-residuals}, we first show the stellar mass of each galaxy (see Table \ref{tab:obs}) against the estimated dominant merger mass (left). $M_{\rm\star,Dom}$\ is not expected to be particularly well correlated with galaxy mass \citep[e.g.,][]{bell2017,dsouzabell2018a}, and we find this to be true for our observed sample. Using dark matter halo catalogs from Illustris-TNG and the abundance matching method, \cite{carlsten2021} found an average relation between the stellar mass of the central galaxy, $\log_{10}M_{\rm \star,Gal}$, and the number of $M_V\,{<}\,{-}9$\ satellites within a 150\,kpc projected radius, $N_{\rm Sat}$\ --- the same definition used in this paper. In Figure \ref{fig:sat-residuals} (center) we show this average relation, and its large estimated scatter, along with the $M_{\rm \star,Gal}$\ and $N_{\rm Sat}$\ for the eight observed galaxies (as well as galaxies from the FIRE simulation, which will be discussed in \S\,\ref{sec:model-comparison}). The stellar masses assumed for this galaxy sample vary relatively substantially throughout the literature. Comparing available methods for stellar mass estimation \citep[e.g.,][]{bell2003,querejeta2015}, we estimate a typical uncertainty of 0.2\,dex on each stellar mass. It is worth noting that the subsequent full possible 0.4\,dex stellar mass range considered for each galaxy is more than half of the full stellar mass range of 0.58\,dex for the sample. This is an unfortunate limitation of stellar mass comparison that cannot be avoided.

To assess the potential impact of galaxy mass on satellite number, we evaluate the residual $N_{\rm Sat}$\ for each galaxy from its predicted average, $\left<N_{\rm Sat}\right>$, estimated from its stellar mass (see Table \ref{tab:obs}) and the \cite{carlsten2021} relation. Given the small dynamic range in stellar mass, yet steep $M_{\rm \star,Gal}{-}N_{\rm Sat}$\ relation, small changes in $M_{\rm \star,Gal}$\ have a large change in expected $N_{\rm Sat}$. It is important to incorporate these uncertainties from the stellar mass of the main galaxy. Accordingly, we run 10,000 Monte Carlo tests to convert the individual points to probability density space. For each test, we assess $M_{\rm \star,Dom}$\ and $N_{\rm Sat}{-}\left<N_{\rm Sat}\right>$\ for each of the eight systems as random draws from a Gaussian random variable with width equal to the estimated uncertainty in each direction. 

We show the results of this Monte Carlo experiment in Figure \ref{fig:sat-residuals} (right). Due to the large overlapping uncertainties, the eight individual measurements naturally cluster into three broad groups in $M_{\rm \star,Dom}$\ --- $8.1\,{\lesssim}\,\log_{10}M_{\rm \star,Dom}\,{\lesssim}\,8.9$, $9\,{\lesssim}\,\log_{10}M_{\rm \star,Dom}\,{\lesssim}\,9.8$, and $10.2\,{\lesssim}\,\log_{10}M_{\rm \star,Dom}\,{\lesssim}\,11$. For each of these three $M_{\rm \star,Dom}$\ bins, we calculate the average $N_{\rm Sat}{-}\left<N_{\rm Sat}\right>$. A clear positive correlation is visible between $M_{\rm \star,Dom}$\ and $N_{\rm Sat}{-}\left<N_{\rm Sat}\right>$\ across the three bins. 

We do note that the distribution of $N_{\rm Sat}{-}\left<N_{\rm Sat}\right>$\ for the eight observed galaxies appears shifted to negative $N_{\rm Sat}{-}\left<N_{\rm Sat}\right>$, relative to the abundance-matched Illustris-TNG distribution. It is possible that this is a statistical effect, though our randomized treatment of the uncertainties should yield a relatively robust distribution. The other possibility is a difference in assumptions between the observations and the abundance matching applied to Illustris-TNG. The most likely of these seems to be a modestly lower stellar mass normalization (likely reflecting a lower stellar mass-to-light ratio) for the halo occupation model used on TNG halos relative to the nearby galaxy sample.

Regardless of the large uncertainties or potential mass-to-light differences, it seems clear that a strong relationship between number of satellites and mass of the dominant merger persists, even after taking into careful account a possible intrinsic relationship with stellar mass of the central.\footnote{In \textsc{appendix} \ref{sec:mdom-mgal} we also consider the relationship defined using the dominant merger ratio, $M_{\rm\star,Dom}/M_{\rm\star,Gal}$, and find no statistical difference.}

\section{Satellite Populations are Assembled During Mergers}
\label{sec:built}

Throughout the previous sections, we have established that a strong relationship exists between the number of satellites around nearby MW-mass systems and the mass of their dominant mergers. Of the eight observed systems considered in this work, two galaxies have estimates of both their first and second most dominant mergers, as well as usable evidence for which of their satellites may have existed prior to the most dominant event --- the MW and M81. Both galaxies are currently experiencing their most dominant merger --- the MW with the LMC, and M81 with M82 --- but their accreted masses (i.e. limit on their previous dominant mergers) have also been measured from their stellar halos (see Table \ref{tab:obs}). Recent work has shown promise in differentiating existing and newcomer satellites in both systems.\footnote{We exclude the very well studied M31 system from this analysis, as its largest merger prior to the dominant merger several Gyr ago that formed its massive halo \citep[e.g.,][see Table \ref{tab:obs}]{dsouzabell2018b} is not known. It is therefore unclear how its largest ancient merger compares to its current merger with the massive satellite M33, and we cannot tell which satellites may have belonged to it.} In this section, we will use this recent work to sketch the evolution of the MW and M81 satellite systems from their past to current dominant mergers, showing that their satellite systems evolve along the relation.

\subsection{Recent MW Satellites}
\label{sec:mw-sat}

It is commonly accepted that the SMC was a satellite of the LMC prior to infall \citep[e.g.,][]{besla2007,besla2012}: at least two of the MW's classical satellites fell in together. However, there has been vigorous discussion about which, if any, of the MW's classical dwarf spheroidals may have previously belonged to the Magellanic system. It is clear that some ultra-faint satellites fell in with the Magellanic Clouds \citep[e.g.,][]{patel2020}, and the motions of two classical MW satellites measured by \textit{Gaia} suggest that they share an orbital plane with the LMC: Fornax and Carina \citep{gaiacollaboration-brown2018,pardy2020}. Modeling the orbits of satellites accreted along with a larger satellite is difficult and highly uncertain \citep[e.g.,][]{dsouzabell2022}, and additional uncertainty comes from differing definitions of former LMC satellites \citep[e.g.,][]{patel2020,pardy2020}. Irrespective of whether Fornax and Carina were accreted with the LMC or not, between 2/7 and 4/7 of the MW’s current classical satellites (within 150\,kpc) were accreted during its merger with the LMC.

In light of these new insights, the star formation histories (SFHs) of satellites in the MW have been analyzed with renewed interest. All of the low-mass satellites of the MW have had their star formation quenched, assumed to be due to ram-pressure stripping of their gas by the MW's circumgalactic medium soon after infall \citep[e.g.,][]{slater&bell2014}. Under this assumption, the SFHs of classical dwarfs may contain information about when they fell into the MW's halo. In an effort to compare the broad SFH properties of the MW's satellites in this way, \cite{weisz2019} used deep \textit{HST}-derived SFHs to calculate two informative quantities: the ages at which each satellite had formed 50\% and 90\% of its stellar mass, $\tau_{50}$\ and $\tau_{90}$. A recent $\tau_{90}$\ (e.g., $\lesssim$2\,Gyr in lookback time), coupled with the absence of young, bright main sequence stars, indicates very recently-halted star formation, while a more ancient $\tau_{90}$\ indicates a `shutdown' of star formation at earlier times. As nearly all low-mass galaxies are anticipated to be star-forming in isolation \citep{geha2012}, $\tau_{90}$\ --- whether indicative of ongoing or recently-halted star formation --- is a promising potential proxy for a satellite's infall time into the central halo \citep[see also][]{dsouzabell2021}. It is worth noting that `pre-processing' of low-mass satellites around massive hosts prior to infall could potentially bias $\tau_{90}$\ to earlier times \citep{jahn2021}. 

\cite{weisz2019} found that the MW's satellites are divided into two groups in $t_{90}$: one with $\tau_{90}\,{>}$\,6\,Gyr and another with $\tau_{90}\,{<}$\,3\,Gyr --- just prior to the infall of the LMC \citep[e.g.,][]{deason2015b}. Among these most recently star-forming satellites are Fornax and Carina, providing further tentative evidence of their possible recent infall into the MW halo. \cite{weisz2019} find similar $\tau_{90}$\ clustering in the satellites of M31, with nearly 50\% exhibiting $\tau_{90}$\ shutdown times between 3--6\,Gyr ago --- just prior to the major merger that is now thought to have formed M31's massive, metal-rich stellar halo \citep{dsouzabell2018b}. This clustering in the star formation shutdown timescale of satellites is a hallmark of the infall of a group with a relatively massive central \citep{dsouzabell2021}.

Given these multiple lines of evidence, for our illustrative analysis in \S\,\ref{sec:mw-m81}, we consider it certain that the LMC and SMC fell in together, and \textit{possible} that Fornax and Carina's infall was associated with the LMC. 

\input{m81}

\subsection{Recent M81 Satellites}
\label{sec:m81-sat}

Though no comparable-quality dynamical evidence exists for the satellites of M81 (and likely won't for the foreseeable future), rich existing resolved star datasets do exist for many M81 satellites, along with corresponding SFH analyses. M82 and NGC 3077 are both commonly assumed to be recent additions to the M81 system, as evidenced by M81's ongoing interaction with both \citep[e.g.,][]{yun1994,okamoto2015,smercina2020}. Additionally, its third most massive satellite, NGC 2976, is thought to have interacted with M81 as recently as $\sim$1\,Gyr ago \citep{williams2010} and therefore is also very likely to be a recent addition on a similar timescale. Given their very comparable infall times, we consider NGC 3077 and NGC 2976 to be past M82 satellites. 

It has been long known that M81's global satellite population exhibits more active and recent star formation than is found in the MW's or M31's satellites \citep{weisz2008}. Using the intuition built from the MW's and M31's satellites, both of which show clustering in their $\tau_{90}$\ timescales, we analyze the broad SFH properties of all 8 available low-mass satellites within 150\,kpc of M81 to assess possible clustering around the infall time of M82. SFHs for each of these satellites were calculated as part of the ANGST survey \citep{weisz2011}. We use these published SFHs to evaluate $\tau_{50}$\ and $\tau_{90}$\ for each satellite and estimate uncertainties from the 16--84th percentile uncertainties on the cumulative SFHs calculated by \cite{weisz2011}. Figure \ref{fig:mw-m81} (right) shows the estimated $\tau_{50}$\ and $\tau_{90}$\ for each of the 8 satellites, color-coded by their $V$-band absolute magnitude, also presented in Table \ref{tab:m81}.

Of these eight satellites, two satellites have ancient $\tau_{90}\,{>}\,9$\,Gyr. The remaining six have $\tau_{90}\,{\leqslant}\,4$\,Gyr, which is comparable to the timescale of M82's infall \citep[e.g.,][]{yun1994}. While the uncertainties on these timescales are large (due to model uncertainty in stellar populations shallower than the main sequence turnoff), three of these satellites --- F8D1, KDG61, and KDG64 --- had very recent star formation ($\tau_{90}\,{\lesssim}\,2$\,Gyr). We assert that these three satellites of M81, along with NGC 3077 and NGC 2976, very likely originated in the M82 system prior to infall into M81's halo. The three satellites with $\tau_{90}$\ between 3--4\,Gyr --- FM1, KK77, and BK3N --- also have quite recent implied infall times, and could be more recent, given their uncertainties. 

Given the existing evidence, we take NGC 3077, NGC 2976, F8D1, KDG61, and KDG64 to be likely former M82 satellites, and FM1, KK77, and BK3N to be possible former M82 satellites.

\subsection{Evolution of the MW and M81 Satellite Systems}
\label{sec:mw-m81}

Estimates exist for the accreted component of both the MW's \citep{bell2008,deason2019,conroy2019,mackerethbovy2020} and M81's \citep{harmsen2017,smercina2020} stellar halos. Therefore, for these two systems, the mass of their largest current satellite ($M_{\rm \star,Dom.\,Sat}$) and total accreted mass ($M_{\rm \star,Acc}$), given in Table \ref{tab:obs}, represent their $M_{\rm \star,Dom}$\ and \textit{previous} $M_{\rm \star,Dom}$, respectively. Given our estimates of the likeliest recent additions to their classical satellite populations presented in the previous sections (\S\,\ref{sec:mw-sat} \&\ \ref{sec:m81-sat}), this provides an opportunity to chart out the evolution of both the MW's and M81's satellite populations with respect to their recent merger history.

\begin{figure*}[t]
\leavevmode
\centering

\begin{minipage}{0.462\textwidth}
    \includegraphics[width={\linewidth}]{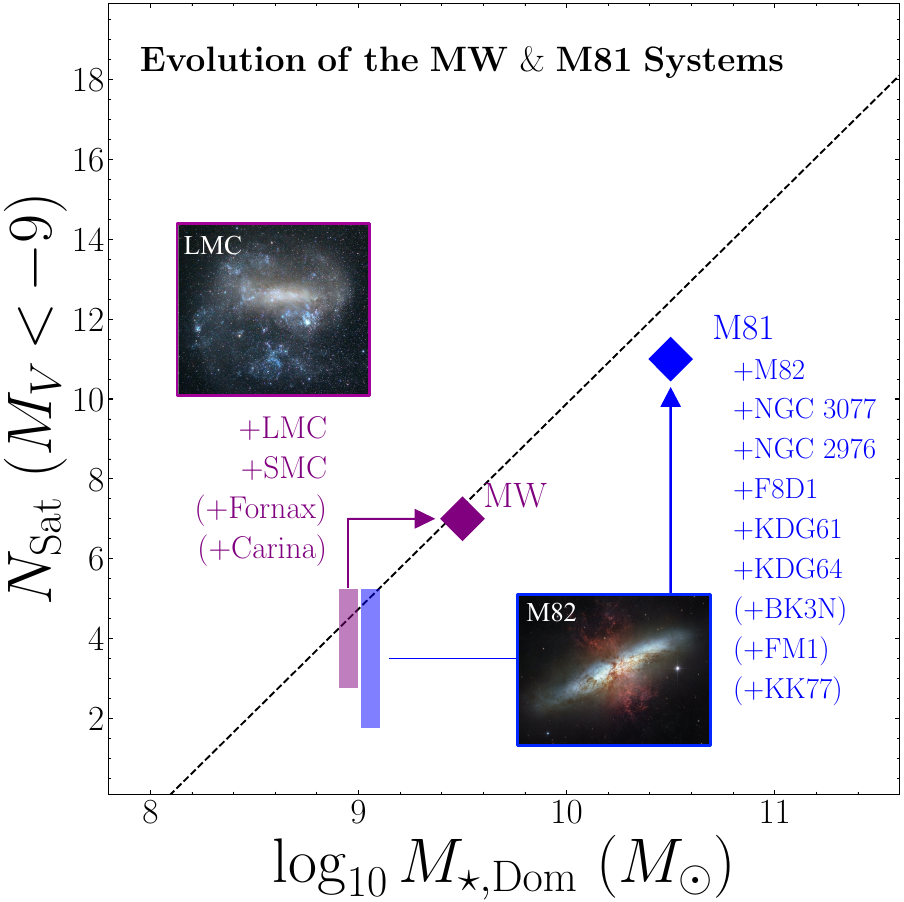}
\end{minipage}
\begin{minipage}{0.513\textwidth}
    \includegraphics[width={\linewidth}]{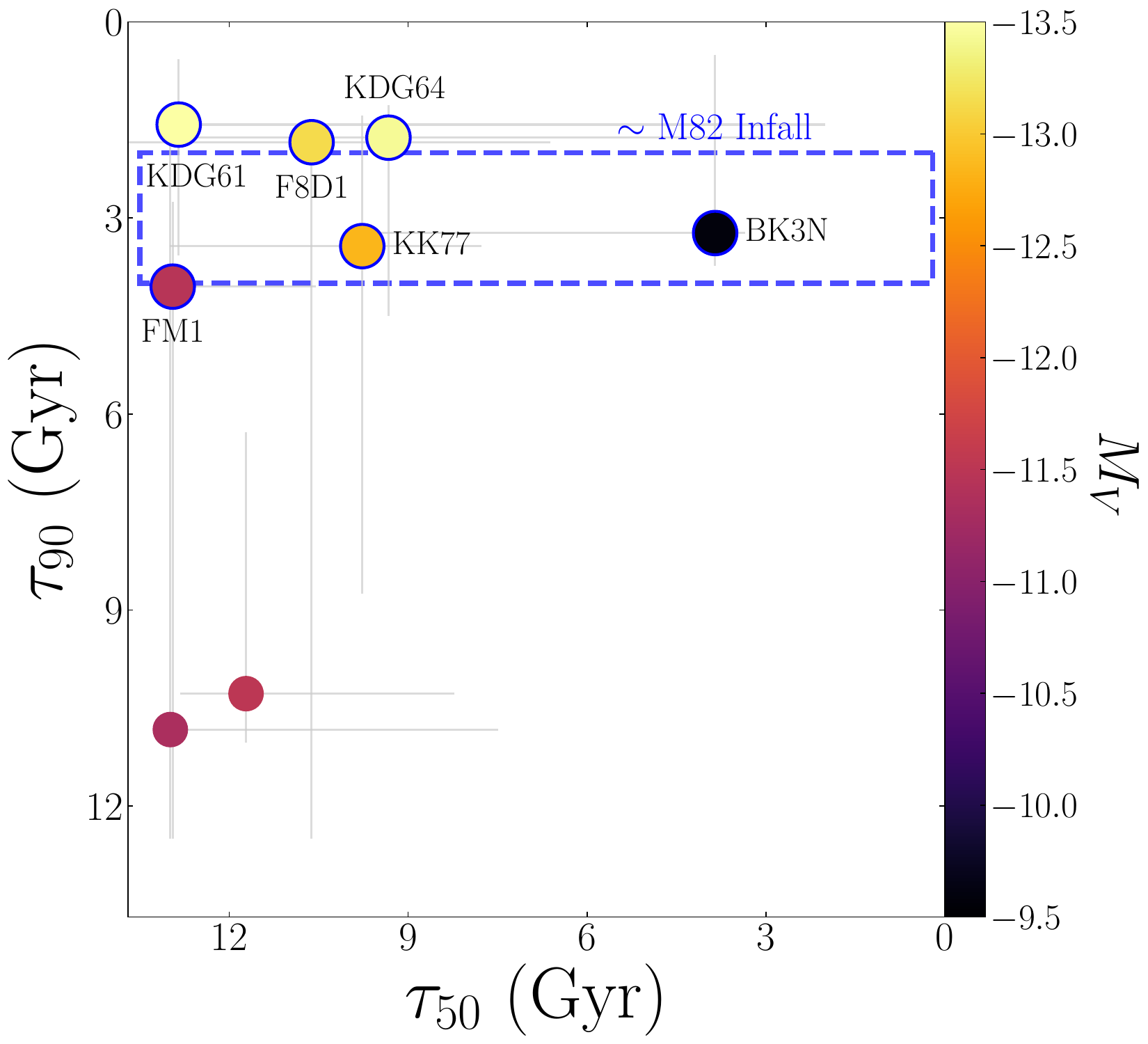}
\end{minipage}

\caption{\uline{Left}: A schematic view of the evolution of the MW's (purple) and M81's (blue) satellite populations along the $M_{\rm \star,Dom}$--$N_{\rm Sat}$\ relation. We show the current position of each system as solid diamonds, assuming their existing satellite populations and dominant mergers (the LMC and M82, images included$^{\textcolor{blue}{1}}$). Next to each galaxy, we list the satellites that we claim in \S\,\ref{sec:mw-sat} \&\ \ref{sec:m81-sat} to have been likely(possibly) accreted during each galaxy's current dominant merger. Finally, we show estimates of the state of the MW and M81 satellite systems prior to their current mergers, represented by their previous $M_{\rm\star,Dom}$\ and a shaded region showing the range of $N_{\rm Sat}$\ at that time, considering the recent addition of both likely and possibly accreted satellites for each. \uline{Right}: The 50\% and 90\% star formation timescales, $\tau_{50}$\ and $\tau_{90}$, for M81's 8 low-mass satellites within a projected radius of 150\,kpc, derived from resolved SFHs \citep{weisz2011}. Galaxies have been color-coded by their absolute $V$-band magnitude. Of these 8, six of these possess $\tau_{90}$\ that falls in the estimated 1--4\,Gyr window of M82's first infall (circled). We consider the three satellites with the most recent $\tau_{90}$\ ($<$2\,Gyr: KDG61, KDG64, and F8D1) to be \textit{likely} former satellites of M82, and the remaining three ($\tau_{90}\,{<}$\,3--4\,Gyr: FM1, KK77, and BK3N) to be \textit{possible} former satellites of M82. \newline $^{\textcolor{blue}{1}}\,${\scriptsize \uline{LMC}: Primoz Cigler, \href{http://astro.primozcigler.net}{http://astro.primozcigler.net}; \uline{M82}: \textit{Hubble Heritage Project}}
}
\label{fig:mw-m81}
\end{figure*}

In Figure \ref{fig:mw-m81} (left), we show the current positions of the MW and M81 relative to the $M_{\rm \star,Dom}$--$N_{\rm Sat}$\ relation, as well as estimates for their positions prior to their respective current dominant mergers with the LMC and M82. While the identified recently-accreted satellites for each are uncertain, both systems tightly track the relation shown in Figs.\ \ref{fig:sat-accretion} \&\ \ref{fig:FIRE-compare} throughout the transition from the previous to current dominant mergers. This suggests that, prior to their current merger state, both galaxies had much sparser satellite populations, comparable to M83, M94, and M101. It is worth noting that, assuming current star formation rates for each galaxy of $\sim$1\ $M_{\odot}\,{\rm yr}^{-1}$, their stellar masses while hosting these sparser satellite populations were $>$96\% of their current values --- further demonstrating the independence of this relationship from stellar mass considerations. The case of M81 is particularly stark, as it appears to have accreted between 55\% and 82\% of its existing ($R\,{<}\,150$\,kpc) classical satellites during its merger with M82. These results suggest that the surviving satellite populations of MW-mass galaxies are assembled primarily during mergers.

\section{Comparison to Simulations}
\label{sec:model-comparison}

Having established in the previous sections that the satellite populations of MW-mass galaxies tightly correlate with the masses of the dominant mergers they have experienced, it is necessary to contextualize this result. How does this empirical relation compare to theoretical predictions?

To first order, CDM predicts that the number of subhalos a central halo hosts should scale with its virial mass. This can be intuited from the much richer satellite populations of massive galaxy clusters, relative to MW-mass groups. However, we are considering MW-mass galaxies, which should occupy halos with a much smaller range of virial masses than the difference between clusters and groups. Particularly in this narrower range of halo mass, where low-mass satellites hold essentially all of the statistical weight, there is considerable precedent for the deviation of satellite galaxy properties from CDM predictions when accounting for the complex impact of baryons (e.g., the `Missing Satellites' and `Too Big to Fail' problems; e.g., \citealt{maccio2010,font2011,brooks2013,wetzel2016}). Given this, it is most fruitful to compare the observed $M_{\rm \star,Dom}$--$N_{\rm Sat}$\ relation against hydrodynamic simulations.

Recent results from the Illustris TNG50 simulation reproduce the predictions from CDM; mainly: a scattered relationship between $N_{\rm Sat}$\ and virial mass \citep{engler2021}. However, \cite{engler2021} also find hints of a correlation between later assembly times and higher satellite number, albeit with high scatter. This may indicate that systems with later-time, larger mergers have more satellites --- qualitatively similar to the trend seen in the observations. However, TNG50 does not resolve the more numerous low-mass classical satellites probed by the observations, making a fair comparison difficult. Ideally, the observations would be compared to a simulation with high enough resolution to capture the entire satellite population.

Currently, the highest-resolution baryonic simulation that meets the necessary criteria to compare complete satellite populations and merger histories in a sample of MW-mass galaxies is the FIRE \citep{hopkins2014,hopkins2018} simulation suite. FIRE has resolved both the satellite galaxy populations \citep[down to $M_{\star}\,{>}\,10^{5}\,M_{\odot}$;][]{garrison-kimmel2019} and stellar halo properties \citep{sanderson2018} of a sample of MW-mass galaxies, in both isolated and Local Group-like environments, with virial masses ranging from $0.8{-}2{\times}10^{12}\,M_{\odot}$.\footnote{See \textsc{appendix} \ref{sec:fire-table}} 

To directly compare FIRE to the eight observed systems, we take the satellite catalogs for each simulated system in FIRE \citep{samuel2020}, assess the number of satellites with $M_{\star}\,{>}\,4{\times}10^5\ M_{\odot}$\ (comparable to $M_{V}\,{<}\,{-}9$; \citealt{smercina2018}) and projected radius $<$\,150\,kpc for 1,000 randomly drawn 3-D orientations, and measure the average $N_{\rm Sat}$\ across these 1,000 different sightlines --- identical to the technique we used to construct the MW's and M31's satellite LFs (\S\,\ref{sec:sat-bg}). We take $M_{\rm \star,Dom}$\ to be the larger of the \textit{peak} stellar mass of each galaxies' largest $z\,{=}\,0$\ satellite or the total accreted mass --- identical to the definition of $M_{\rm \star,Dom}$\ for the observed sample. We provide accreted masses, dominant satellite masses, and sightline-averaged satellite counts for the FIRE galaxies in Table \ref{tab:fire} in \textsc{appendix} \ref{sec:fire-table}. 

We combine the Local Group-analog `ELVIS on FIRE' (treated as individual systems) and isolated `m12' suites, as our observational comparison set combines a number of different environments.\footnote{The ELVIS suite was run at twice the resolution of the m12 suite, but with identical physics. See \textsc{appendix} \ref{sec:fire-table}.} Following \S\,\ref{sec:significance}, we use \texttt{emcee} to calculate the posterior probability distributions for the parameters of the model defined in Equation \ref{eq:1} given this mock-observed FIRE sample. We show the results of this analysis in Figure \ref{fig:FIRE-compare}, where the results for the FIRE simulation are shown in orange. As with the observations (\S\,\ref{sec:significance}), we show the model with highest likelihood of describing the FIRE systems and the 16--84\% confidence region, as well as the marginalized single and joint posterior distributions for the model parameters (right panel). The difference between the FIRE and observed samples is striking. Unlike the observed systems, FIRE shows shows no correlation and large intrinsic scatter at a given $M_{\rm \star,Dom}$. In \textsc{appendix} \ref{sec:fire-auriga} we compare FIRE to the Auriga simulation to show that this result is, in general, independent of the differences between them. 

We also show the FIRE systems in Figure \ref{fig:sat-residuals} for comparison, with stellar masses taken from \cite{samuel2020} and \cite{santistevan2021} (see Table \ref{tab:fire}). Surprisingly, the FIRE systems exhibit a positive correlation between $M_{\rm \star,Dom}$\ and $M_{\rm \star,Gal}$, in contrast with the observations and expectations \cite[e.g.,][]{bell2017}.\footnote{See also \textsc{appendix} \ref{sec:mdom-mgal}.} Conversely, FIRE does does not exhibit any discernible correlation between $N_{\rm Sat}$\ and $M_{\rm \star,Gal}$. This lack of a trend between  $M_{\rm \star,Gal}$\ and $N_{\rm Sat}$\ in FIRE has been previously noted and is thought to be related to the smaller range of halo masses in FIRE (0.9--1.7$\times$10$^{12}\ M_{\odot}$) than e.g., the \cite{carlsten2021} analysis (0.8--8$\times$10$^{12}\ M_{\odot}$), which may allow the more effective tidal disruption from more massive central galaxies to dominate any scaling of satellite number with virial mass \citep{samuel2020}. We will discuss this further in \S\,\ref{sec:tidal}. The FIRE $N_{\rm Sat}$\ residuals exhibit essentially pure scatter in $N_{\rm Sat}{-}\left<N_{\rm Sat}\right>$\ for a fixed $M_{\rm \star,Dom}$, which in turn is entirely consistent with the TNG-abundance matching results. It is, however, completely unlike the correlation visible in the observed sample, despite a very comparable stellar mass range as seen in the left panel. We observe no statistical difference between the LG-like ELVIS hosts and the isolated m12 hosts. In addition, as M104 is also more massive (by a factor of $\sim$2 in $M_{\rm \star,Gal}$) than the range of galaxies produced in FIRE, we also conduct the comparison analysis between the observed sample excluding M104 and FIRE, and find identical results.

It is important to note that FIRE produces only half the logarithmic dynamic range in dominant merger mass experienced by MW-mass galaxies compared to the observations --- $\log_{10}M_{\odot}\,{\sim}\,9{-}10.4$\ vs. $\log_{10}M_{\odot}\,{\sim}\,8.3{-}11$. For reference, this lack of diversity is borne out in the comparable-resolution Auriga simulation as well (see \textsc{appendix} \ref{sec:fire-auriga}), though neither simulation imposes a selection constraint on the merger history. It is unclear whether this lack of diversity is related somehow to the by-design choice of specific halos in these cosmological zoom-in simulations thought to be capable of hosting a MW-mass disk-like galaxy (as cosmological simulations typically show a broader range more consistent with the observations; \citealt{deason2015b}, \citealt{dsouzabell2018a}), or possibly indicates a more fundamental problem with halo occupation. An important first step towards better understanding this tension will likely be to work towards producing comparable-resolution simulations that encompass a larger range of halo masses. 

This being said, it is entirely clear that even if FIRE were to encompass a broader range of halo masses, and were it shown that such a broader halo mass range is more consistent with the observed galaxy sample considered here, the considerable scatter in $N_{\rm Sat}$\ at a fixed $M_{\rm \star,Dom}$\ (see Figure \ref{fig:FIRE-compare}) would persist. This large scatter is present in other baryonic simulations (e.g., TNG50 and Auriga; \citealt{engler2021}, \textsc{appendix} \ref{sec:fire-auriga}), regardless of the halo mass range used. This suggests that, while there may well be an inevitable scaling of satellite number with virial mass, the tight empirical $M_{\rm \star,Dom}$--$N_{\rm Sat}$\ relation is not forthcoming in current galaxy formation models.

This brings us to a last, important consideration: despite the lack of a tight $M_{\rm \star,Dom}$--$N_{\rm Sat}$\ relation, the satellite count normalization (given by $N_{9.5}$\ in our adopted model) is almost identical in both FIRE and the observations. Rather than indicative of specific calibration on the part of the models to match the MW, we contend that this reflects the status of the MW's satellites as a long-standing benchmark for the community. This suggests that benchmarking against the MW, irrespective of merger history, has allowed high resolution zoom-in simulations to relatively faithfully reproduce the MW's satellites while failing to fully capture the drivers of the scaling relation between MW-mass galaxies' satellite number and the mass of the their dominant merger.

\section{Implications for Galaxy Formation}
\label{sec:implications}

Our direct comparison of the satellite populations and dominant mergers of eight nearby systems benefited from impressive observational progress on both the satellite and stellar halo fronts. The next decade will undoubtedly see this sample of galaxies grow, encompassing systems in different environments and with different merger histories. These future observations will allow us to better understand and interpret the observational findings of this paper; mainly, the empirical relation between satellite number and dominant merger mass. Acknowledging the likely future refinement of our results, we aim to evaluate the scientific insight gained from the existence of this relation, and discuss what we find to be a number of important implications for galaxy formation.

\subsection{Sketching the Problem}
It is clear from the results of \S\,\ref{sec:model-comparison} that current galaxy formation models have used the MW's satellite population as a benchmark for the successful modeling of low-mass galaxy formation --- \textit{irrespective} of group merger history. The observations, however, tell a different story. In nature, even if a broad overall trend exists between virial mass and satellite number, there are physical processes that drive the emergence of a relationship between a MW-mass galaxy's satellite population and its merger history. These processes must be fundamental to how the universe forms low-mass galaxies, and yet are poorly represented in, or absent from, current galaxy formation simulations.  

The first important realization is that, while the infall of satellites during galaxy mergers may appear, on the surface, to be a natural consequence of hierarchical galaxy formation (as discussed in the Introduction), the empirical relation between $M_{\rm \star,Dom}$\ and $N_{\rm Sat}$\ does not emerge from CDM alone. As discussed previously in both the Introduction and \S\,\ref{sec:model-comparison}, to zero order, the number of subhalos near a MW-mass dark matter halo is set by the MW-mass halo's mass --- more mass, more subhalos --- the motivation for halo occupation modeling efforts \citep[e.g.,][]{carlsten2021}. The merger of a MW-mass halo with a large satellite and its group would therefore ultimately be expected to have little effect on the number of subhalos and the satellites in them. Those subhalos would have inevitably fallen in at some stage during the growth of the MW-mass halo; the infall of those subhalos in a group simply clusters their accretion tightly in time. This is what is indeed borne out in high-resolution DMO simulations: a weak or nonexistent correlation between subhalo number and the mass of the most massive accretion, but strong clustering in accretion time around the accretion times of the most massive progenitor satellites \citep{dsouzabell2021}. The emergence of a relation between present-day satellite populations and merger history, clearly visible despite the expected importance of a scaling with overall halo mass, therefore cannot be a direct consequence of hierarchical structure formation, but instead must be due to more complex galaxy physics. 

Yet, a similarly tight relation fails to emerge even when baryonic physics are introduced, ranging from semi-analytic model-based analyses, \citep[e.g.,][]{bose2020} to cosmological hydrodynamic simulations \citep{engler2021}, and high-resolution baryonic zoom-in simulations (\S\,\ref{sec:model-comparison}). These model results are all in agreement with the expectations from hierarchical structure formation, yielding a highly-scattered trend between satellite number and virial mass, yet produce at worst no trend, or at best a highly-scattered trend, between satellite number and merger history. Though some of these simulations, by design of their focus on MW-mass galaxies, feature a somewhat limited halo mass range \citep[see ][for a summary of halo mass ranges in current high-resolution simulations]{engler2021}, as an ensemble they fail to capture the tightness of the empirical correlation, regardless of the presence of an intrinsically scattered trend driven by virial mass. The contrast between the highly scattered trend of $N_{\rm Sat}$\ with virial mass in CDM, and the tight empirical trend shown here, indicates that the stellar mass of the dominant merger is an independent driver of satellite number at the MW-mass scale. 

One clear but uncomfortable implication of the empirical $M_{\rm \star,Dom}$--$N_{\rm Sat}$\ relation is that the halos of MW-mass galaxies which have never experienced a massive merger possess only sparse satellite populations of $M_V<-9$ satellites within 150\,kpc. Groups like M94's or M83's have a comparable number of such $R_{\rm proj}<150$\,kpc satellites to the 1--3 likely LMC satellites discussed above as well as model predictions for the LMC \citep[e.g.,][]{dooley2017b}, and potentially fewer than a group like M82's. These groups only gain a large number of satellites by merging with a smaller group, as is clear from M81 in the discussion above, or from the distribution of star formation quenching times of the M31 satellites, suggesting many of them fell in during M31's most massive merger event \citep{weisz2019,dsouzabell2021}. It is no wonder that galaxy formation models do not reproduce such behavior: in the case of an almost completely quiescent merger history, the baseline satellite population of a MW-mass galaxy is quite sparse, similar to M94. Without fully understanding the role of the dominant merger in setting satellite number, this could again place the satellites of MW-like galaxies in tension with the expectation of DMO and galaxy formation models, harkening back to the early days of the `Missing Satellites Problem' \citep{klypin1999,moore1999}. This is a problem that requires explanation, and we suggest several possible avenues for further investigation in the context of galaxy formation physics. 

\subsection{Virial Radius}
An important consideration is the relative virial radii of the central's and dominant satellite's halos, which amounts to a possible `aperture effect'. Bigger halos have larger virial radii; therefore, a smaller fraction of a MW-mass group's satellites may reside within 150\,kpc than in an LMC-mass halo. \cite{dsouzabell2021} note that time-dependent variation in satellite radial profiles occurs as satellite populations `virialize' during a massive merger, and \cite{samuel2020} note similar variability in satellite populations in the FIRE simulations, due to the shorter timescales of satellites at pericenter. These studies demonstrate that it is imperative to consider the impact of radial selection both observationally and theoretically. In our analysis, we have been careful to treat the observations and models consistently, but acknowledge that future (especially observational) efforts with access to faint satellites over much larger volumes would do much to reduce this source of uncertainty. With that said, such a selection effect would not entirely solve the problem of an M94-like sparse satellite population representing the `baseline' satellite population of a MW-mass group. 

\subsection{Enhanced Tidal Disruption}
\label{sec:tidal}
It is possible that the number of satellites correlates with late accretion because early-forming satellites of the primary galaxy are destroyed more efficiently by tides than current models suggest. Models of MW-like accretion histories indicate that most all luminous satellites ever accreted by a MW-mass galaxy should be disrupted by present-day \citep[e.g., roughly 90\% of satellites are disrupted in][]{bullock&johnston2005}. However, the tidal disruption of satellites is highly sensitive to a number of difficult-to-model physical properties, including the shape of the gravitational potential in the presence of a stellar disk \citep{donghia2010,garrison-kimmel2017b} and the density structure of satellites \citep{Pena2010,brooks2014}, which is in-turn highly sensitive to prescriptions of stellar feedback. Because tidal disruption is so effective, modest changes in the modeling of its effectiveness may dramatically affect the number of surviving satellites. 

An excellent example of this is the lack of a relation between $M_{\rm \star,Gal}$\ and $N_{\rm Sat}$\ in FIRE \citep{samuel2020}. It seems as though, when considering a narrow enough range in halo mass (a factor of two, in the case of FIRE), changes in tidal disruption efficiency, driven by scatter in properties of the host galaxy's disk, may dominate over the impact of halo mass on satellite number. While this does not specifically foreshadow a solution to the $M_{\rm \star,Dom}$--$N_{\rm Sat}$\ relation, it does highlight that the efficiency of tidal disruption may be variable from galaxy-to-galaxy and therefore may be an important process for consideration. 

\subsection{Group-Scale Feedback}
It is also possible that the formation of visible satellites could be affected by a large central galaxy through group-scale feedback, from reionization or other baryonic effects. The reionization of the universe is an important ingredient of the current \lcdm\ galaxy formation framework in shaping the properties of low-mass galaxies. Though debate over the dominant source of ionizing photons continues, it has long been suggested that reionization should suppress star formation in low-mass dark matter halos, and therefore in the lowest-mass dwarf galaxies \citep[e.g.,][]{bullock2000}. This idea appears to be confirmed by the uniformly ancient stellar populations in ultra-faint dwarf galaxies around the MW \citep[e.g.,][]{brown2014}. Although reionization is typically imposed in models as a relatively uniform `event', due in large part to the complex sub-grid physics involved with the production and escape of ionizing photons from early galaxies \citep[e.g.,][]{ma2015}, it is generally agreed that reionization was likely a localized process, with variations based on localized star formation conditions, that could have affected the formation of very low-mass galaxies \citep{benson2003,busha2010,lunnan2012,ocvirk2013}. 

Such `patchy' reionization, or other early-time group-scale feedback processes, could be an interesting possiblity to explain the intrinsic dearth of satellites near MW-mass galaxies, relative to lower mass galaxies like the LMC. For example, \cite{tang2009} suggest that concentrated bursts of star formation in proto-MW-mass galaxies, coincident with early bulge formation, could substantially elevate the temperature of the circumgalactic halo gas to beyond the virial radius. Additionally, recent comparisons of the EAGLE and Auriga simulations show that, depending on the adopted prescriptions, MW-mass galaxies may inject substantial feedback from active galactic nuclei on group-scales at early times --- even beyond the virial radius \citep{kelly2021}. It is possible that such group-scale feedback could suppress the growth of even `classical' satellites in the central galaxy's immediate vicinity. 

This group-scale feedback would be expected to scale with the properties and formation history of the central galaxy, likely being more intense for more massive centrals with more intense early star formation. Group-scale feedback may also act in concert with other processes, by affecting the properties of the satellites or their radial distribution, therefore changing their susceptibility to tidal destruction, creating collective effects that are larger than those expected from each process acting alone.

\subsection{Time of Accretion}
When discussing tidal disruption, and the impact external feedback processes may have on its efficiency, it is important to consider the effect of time. Time has already been discussed as an important player in the modulation of satellites' radial distribution, but it is also a fundamental parameter in $M_{\rm \star,Dom}$\ itself. Consider: all but two of the dominant mergers in the eight nearby galaxies are very recent, either occurring right now (MW, M81, M101) or occurred in the last 3\,Gyr \citep[M31, Cen A, M104;][]{dsouzabell2018b,wang2020,cohen2020}. This is in large part due to the continued growth of galaxies through cosmic time --- by the present day, galaxies are more massive and therefore the `time of accretion' for the most massive satellites is intrinsically more likely to be recent \citep[e.g.,][]{dsouzabell2018a}. This is particularly well-demonstrated by recent evidence for a massive early MW accretion that was heavily dark-matter dominated \citep[the \textit{Gaia}-Enceladus-Sausage event][]{helmi2018,belokurov2018}. Such a merger certainly does not qualify as the MW's dominant merger in the definition of $M_{\rm \star,Dom}$\ used in this paper, but may have represented a comparable-mass `ecosystem' merger. In this context, the observed $M_{\rm \star,Dom}$--$N_{\rm Sat}$\ relation could actually be thought of as a relationship between MW-like galaxies' current satellite populations and their \textit{late-time} accretion history. In the context of tidal destruction, for example, this allows long timescales for preexisting satellites to be destroyed prior to the recent dominant merger.

It is important to note that DMO simulations have long predicted that the abundance of subhalos in a central halo is related to the formation history of the central --- halos that form earlier should host fewer present-day subhalos \citep[e.g.,][]{ishiyama2009}. This behavior is also captured in high-resolution baryonic simulations \citep[e.g., TNG50;][]{engler2021}. It has similarly been suggested that subhalo abundance may scale with host concentration, which in turn correlates with time \citep{mao2015}. While these correlations do not directly explain the tight $M_{\rm \star,Dom}$--$N_{\rm Sat}$\ relation found in this paper, they may highlight the importance of an underlying relationship between $M_{\rm \star,Dom}$\ and the time of accretion.

\subsection{Moving forward}
Confirming the relative importance of these, or other, mechanisms as drivers of this relation will require a concerted observational effort on several fronts. The first of these is building a larger sample of galaxies for which both $M_{\rm \star,Dom}$\ and $N_{\rm Sat}$\ have been measured, and search for any outliers or unique stages of the merger history (see \textsc{appendix} \ref{sec:m51-ngc4565}). This will require large samples of galaxies with resolved stellar halo properties from \textit{HST}, \textit{JWST}, the \textit{Nancy Grace Roman Space Telescope} and ground-based facilities, as well as well-measured satellite populations, including either resolved stellar populations or SBF distances. In order to circumvent the potential aperture effects induced by a 150\,kpc radial cut, it is necessary to expand our already wide-field surveys out to at least the virial radius of the host galaxies ($\sim$250--300\,kpc), and ideally well beyond given evidence that many `associated' satellites should lie beyond the virial radius \citep{bakels2021,dsouzabell2021}. Instruments such as Subaru HSC and the Vera Rubin Observatory will crucial to this effort. 

Finally, detailed measurements of the SFHs and LOS velocities of satellites will provide much-needed distinguishing power. As discussed in \S\,\ref{sec:built}, the SFHs of satellites, particularly in systems with ongoing mergers, will provide important clues about the distribution of infall times, relative to the merger. The importance of considering the SFHs of satellites in the context of the $M_{\rm \star,Dom}$--$N_{\rm Sat}$\ could already be reflected in the claimed low quenched fractions of satellites in the Local Volume \citep{bennet2019,karunakaran2021}. Additionally, velocity measurements may help to understand if it is possible that a large influx of satellites during a recent merger may help to explain the perplexing `planes of satellites' observed around galaxies like the MW, M31, and Cen A \citep[e.g.,][]{metz2008,pawlowski2012,ibata2013,muller2018b,samuel2021}.

In short, this observed relationship between the mass of the most dominant merger and number of satellites in MW-mass galaxies may require a fundamental shift in our thinking about the build-up of satellite populations and, consequently, low-mass galaxy formation. The tension with models indicates that this correlation goes beyond the intuitive consequences of hierarchical galaxy formation, and may require consideration of new or updated physical ingredients in our models. Larger galaxy samples and deeper investigations into the SFHs and motions of satellites in nearby galaxy groups will guide models and help us to better understand the emergence of this relation. 

\section{Conclusions}
\label{sec:conclusions}

Motivated by the infall of MW satellites associated with the LMC and the broad diversity in the satellite populations of nearby MW-mass galaxies, throughout this paper we have, for the first time, investigated a possible relationship between these diverse satellite populations and their equally diverse merger histories as traced by their most dominant mergers. For this effort, we define a new metric for the mass of a galaxies' most dominant merger, $M_{\rm \star,Dom}$, as the larger of either its total accreted mass $M_{\rm \star,Acc}$, or the mass of its most massive current satellite, $M_{\rm \star,Dom.\,Sat}$. Using this new definition of $M_{\rm \star,Dom}$, and the satellite populations of eight nearby galaxies, we find:
\begin{enumerate}[topsep=5pt,itemsep=0pt,left=2pt]
    \item We find a strong observed correlation between $M_{\rm \star,Dom}$\ and the number of $M_{V}\,{<}\,{-}9$\ satellites within a projected 150\,kpc ($N_{\rm Sat}$) of the eight nearby MW-mass galaxies in which these two properties have been best measured. 
    \item Assuming a simple linear model between $\log_{10}M_{\rm\star,Dom}$\ and $N_{\rm Sat}$, in concert with MCMC sampling methods, we find an unexpectedly tight relation, which remains even after accounting for possible intrinsic relationships with galaxy stellar mass. We conclude that the number of satellites around a MW-mass galaxy is a strong function of its largest merger partner.
    \item Using recent evidence from orbital and SFH modeling, we empirically demonstrate that both the MW's and M81's satellite populations have evolved along the observed $M_{\rm\star,Dom}$--$N_{\rm Sat}$ relation over the course of their current dominant mergers with the LMC and M82, respectively. Consequently, we estimate that \textit{more than 50\%} of M81's current satellites within a projected 150\,kpc were accreted during its recent merger with M82. 
    \item We compare the observed relation to theoretical MW-like systems from the FIRE simulation, assessing average satellite populations within projected radii $<$150\,kpc across many random orientations. We find the simulated galaxies' behavior to be completely different from the observations. Instead of the tight observed relationship, FIRE exhibits no discernible relationship between $M_{\rm\star,Dom}$\ and $N_{\rm Sat}$, with a large intrinsic scatter in $N_{\rm Sat}$\ at a fixed $M_{\rm \star,Dom}$, but does produce the correct number of satellites for a MW-like $M_{\rm \star,Dom}$. Given the similarity between the results from DMO simulations, semi-analytic models, and these high-resolution hydrodynamic simulations, this tension suggests that, while current simulations can reproduce the small-scale structure observed around the MW, they are not fully capturing the diversity of merger histories essential to understanding the formation of other MW-mass galaxies and their satellites.
    \item The $M_{\rm \star,Dom}$--$N_{\rm Sat}$\ relation directly implies that galaxies like M94 or M83, with difficult-to-explain sparse satellite populations, may represent the `baseline' present-day satellite population of a MW-mass galaxy, in the absence of any larger merger throughout its life.
    \item We speculate that some combination of heavily time-dependent satellite radial distributions, enhanced tidal disruption, and early-time group-scale feedback in the vicinity of a massive central may help to explain the emergence of the $M_{\rm \star,Dom}$--$N_{\rm Sat}$\ relation. The `time of accretion' may be an important factor to consider when exploring the validity of these different physical mechanisms.
\end{enumerate}

$M_{\rm \star,Dom}$\ and $N_{\rm Sat}$\ are proxies for two of the most fundamental aspects of hierarchical galaxy formation theory --- galaxy merger histories and the buildup of satellite galaxy populations. Though the infall of satellites during merger events may feel intuitive, the tight scaling relation described here \textit{cannot} be explained as a natural consequence of traditional hierarchical structure formation. The solution must be sought in the context of galaxy formation physics, in which any trend is currently either highly scattered or nonexistent. While future observational facilities will deepen our understanding of the relationship between these proxies, as it currently stands these results, which demonstrate a surprisingly acute difference between their predicted and observed connection, constitute an important tension that must be understood and addressed if we are to achieve a holistic understanding of galaxy formation. The ability of models to reproduce this relationship, including the general diversity in the observed merger histories of MW-like galaxies, may serve as a guide for successfully coupling small- and large-scale galaxy physics, including stellar feedback and environmental processes. \\

We thank the anonymous referee for a careful and thoughtful review that improved this paper. We also thank Andrew Wetzel, Monica Valluri, Jeremy Bailin, Mario Mateo, JD Smith, Julianne Dalcanton, Oleg Gnedin, and Ian Roederer for valuable discussions and feedback. 

A.S.\ was supported by NASA through grant \#GO-14610 from the Space Telescope Science Institute, which is operated by AURA, Inc., under NASA contract NAS 5-26555. E.F.B.\ was partly supported by the National Science Foundation through grant 2007065 and by the WFIRST Infrared Nearby Galaxies Survey (WINGS) collaboration through NASA grant NNG16PJ28C through subcontract from the University of Washington.

This research uses cosmological zoom-in baryonic simulations of Milky Way-mass galaxies from the Feedback In Realistic Environments (FIRE) simulation project, run using the Gizmo gravity plus hydrodynamics code in meshless finite-mass (MFM) mode \citep{hopkins2015} and the FIRE-2 physics model \citep{hopkins2018}. We thank the FIRE team for helping to provide us with electronic versions of the FIRE satellite catalogs, and for valuable advice on working with the data. 

This research has made use of the following astronomical databases and catalogs: the SIMBAD database, operated at CDS, Strasbourg, France \citep{wenger2000}. The VizieR catalogue access tool, CDS, Strasbourg, France (DOI: 10.26093/cds/vizier). The original description of the VizieR service was published in \cite{vizier}. The Extragalactic Distance Database \citep{tully2009}. 

\software{\texttt{Matplotlib} \citep{matplotlib}, \texttt{NumPy} \citep{numpy-guide,numpy}, \texttt{Astropy} \citep{astropy}, \texttt{SciPy} \citep{scipy}, \texttt{emcee} \citep{foreman-mackey2013}, \texttt{corner} \citep{corner}}

\vspace{12pt}
\bibliographystyle{aasjournal}

\appendix 

\section{Limits on the satellite populations of NGC 4565 and M51}
\label{sec:m51-ngc4565}
\begin{figure}[t]
\leavevmode
\centering
\includegraphics[width={0.75\linewidth}]{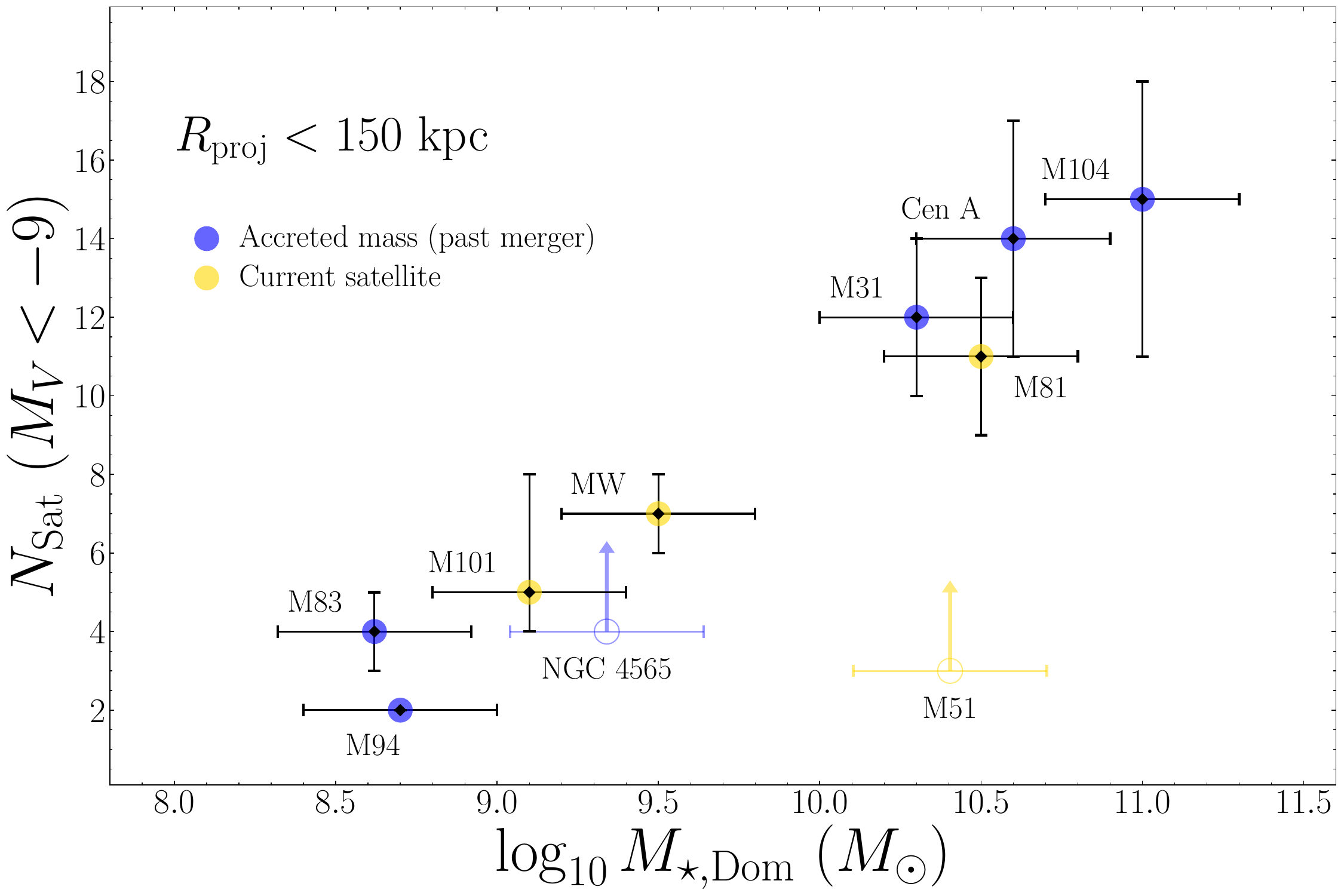}
\caption{Same as Figure \ref{fig:sat-accretion}, but also showing NGC 4565 and M51 (transparent open circles), for which we have assessed lower limits on $N_{\rm Sat}$.}
\label{fig:m51-ngc4565}
\end{figure}

In addition to the eight galaxies considered in this paper, estimates of the satellite populations exist for two other galaxies with well-measured $M_{\rm \star,Dom}$: NGC 4565 and M51. NGC 4565 has an estimated accreted mass from the GHOSTS survey \citep{harmsen2017} which is substantially more massive than any existing satellite. M51 is currently interacting with its well-studied massive satellite, M51b --- a similar mass to M81's dominant satellite, M82. It is unlikely that M51's current accreted mass exceeds M51b's stellar mass, so M51b's mass can be assumed to represent $M_{\rm \star,Dom}$. Both galaxies' satellite populations were studied by \cite{carlsten2021}, but they have large uncertainties stemming from inconclusive SBF distance measurements, similar to M104. However, both have significantly fewer confirmed satellites than M104, and much larger uncertainties on the faint-end of their LFs. Therefore, we treat them as lower limits, compared to the rest of the sample. Figure \ref{fig:m51-ngc4565} shows $M_{\rm \star,Dom}$\ plotted against $N_{\rm Sat}$, identical to Figure \ref{fig:sat-accretion}, but with the lower limits on $N_{\rm Sat}$\ for NGC 4565 and M51 included. 

$N_{\rm Sat}$\ limits for both NGC 4565 and M51 are consistent with the the relation, though M51 lies particularly low relative to the other galaxies with its dominant merger mass (e.g., M31 and M81). If M51's census were complete, it would be a strong outlier from the $N_{\rm Sat}$--$M_{\rm \star,Dom}$ relation, and will be important to understand in detail in future efforts. In light of this, we consider two possible factors to be particularly important to reflect upon and investigate in the future.

The first, as mentioned above, is that the census of M51's satellite population is incomplete, and possibly \textit{highly} incomplete due to its merger stage. \cite{watkins2015} showed that M51's interaction with M51b has distributed tidal debris to radii of at least a projected 50\,kpc from its center ($<$100\,kpc in 3-D radius). This debris has a surface brightness of $\mu_{V}\,{<}\,26$\,\magsqarc, which is brighter than the majority of the faint classical satellites in the Local Group \citep{mcconnachie2012}. \cite{carlsten2020} show that at least half of M51's possible satellites lie in this region dominated by bright tidal debris, and other recent studies of the radial distribution of classical satellites around MW-like hosts also suggest that this projected radial range (i.e.\ $\lesssim$50\,kpc) should encompass a substantial fraction of the central's satellite population \citep[e.g.,][]{samuel2020}. It is entirely possible that this bright \textit{in situ} contamination uniquely impacts the completeness of current satellite searches around M51, relative to other galaxies with much lower surface brightness accreted halos.  

A second possibility is that M51 is truly deficient in satellites within its central 150\,kpc, due to the current stage of its merger with M51b. \cite{dsouzabell2021} show that during a merger between a MW-mass galaxy and a massive satellite, the orbit of the massive satellite influences the bulk distribution of the satellite population. When the massive satellite is close to its first pericenter, its satellites are also close to pericenter and are unusually centrally concentrated (e.g., the Milky Way's; see \citealt{samuel2020}). At later merger stages, the newly-arrived satellites tend to be towards their apocenters (for a timescale of a few Gyrs), which often lie well outside the virial radius. As M51's merger with M51b is in a relatively late stage, it is very possible that it is temporarily deficient in satellites within its inner 150\,kpc. If this is the case, one would expect a significant number of satellites at substantially larger distance from M51 --- a hypothesis that is straightforward to test with future observations.

Independent of the importance of either of these two factors, it is clear that M51's satellite population will be a critical case study to better understand the emergence of the $M_{\rm \star,Dom}$--$N_{\rm Sat}$\ relation. 

\section{Dominant Merger Ratio, $M_{\rm\star,Dom}/M_{\rm\star,Gal}$, vs.\ $N_{\rm Sat}$}
\label{sec:mdom-mgal}

\begin{figure}[!h]
\leavevmode
\centering

\begin{minipage}{0.51\linewidth}
    \includegraphics[width={\linewidth}]{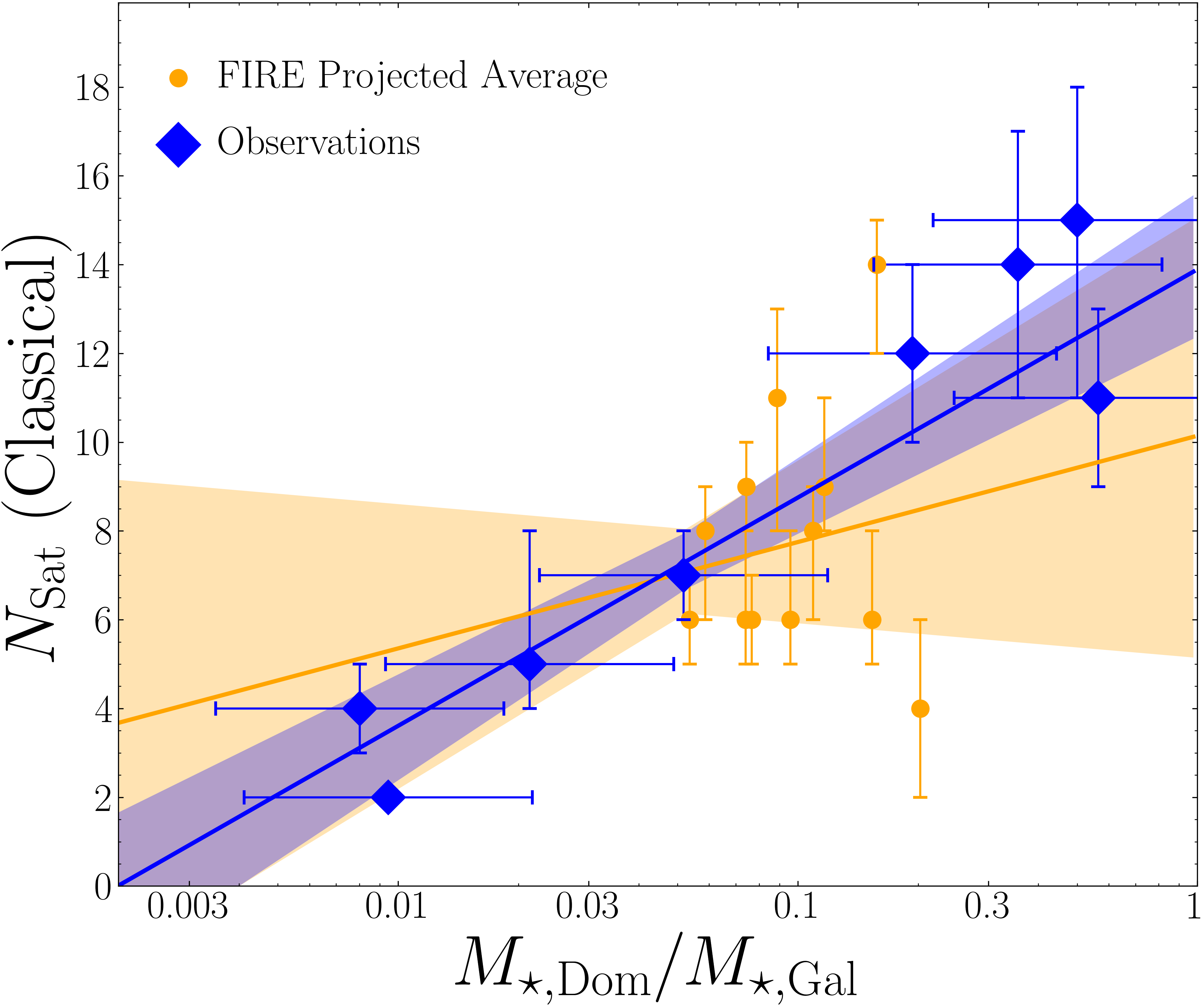}
\end{minipage}
\begin{minipage}{0.45\linewidth}
    \includegraphics[width={\linewidth}]{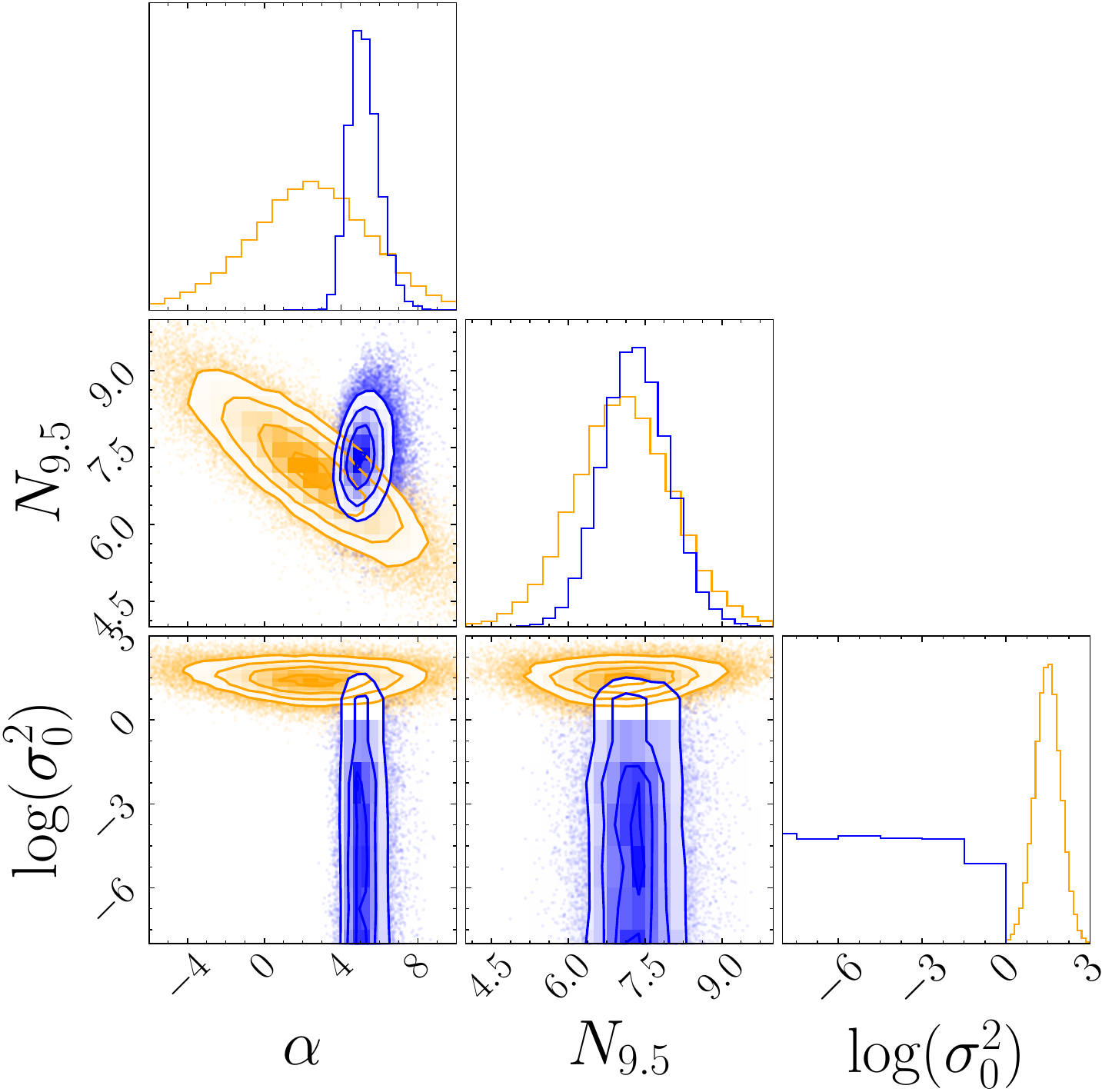}
\end{minipage}

\caption{Identical to Figure \ref{fig:FIRE-compare}, but with $M_{\rm\star,Dom}/M_{\rm\star,Gal}$\ along the $x$-axis.}
\label{fig:FIRE-compare_mdom-mgal}
\end{figure}

Here we assess the impact of normalizing $M_{\rm\star,Dom}$\ to $M_{\rm\star,Gal}$\ --- a `dominant merger ratio'. We adopt a slightly modified version of Equation \ref{eq:1}:
\begin{equation}
        N(\alpha,N_{9.5};M_{\rm \star,Dom}) = \alpha (\log_{10}{M_{\rm \star,Dom}} - 9.5 + \log_{10}{M_{\rm \star,Gal}}) + N_{9.5}.
\end{equation}
We repeat our MCMC anlaysis from \S\,\ref{sec:model-define}, using this updated model. Figure \ref{fig:FIRE-compare_mdom-mgal} shows the results of this analysis. While uncertainties on $M_{\rm\star,Dom}/M_{\rm\star,Gal}$\ are higher than for $M_{\rm\star,Dom}$\ alone, we find results that are very comparable with those presented in Figure \ref{fig:FIRE-compare}: a tight positive relation, with intrinsic scatter consistent with zero.

We also repeat this analysis of the dominant merger ratio with the FIRE simulations. We find a more significant correlation between $M_{\rm\star,Dom}/M_{\rm\star,Gal}$\ and $N_{\rm Sat}$\ than $M_{\rm\star,Dom}$\ alone: while the distribution of slopes is very broad, the average slope is positive. However, the intrinsic scatter is still far higher than the observed relation. The increased correlation observed in Figure \ref{fig:FIRE-compare_mdom-mgal} could be driven at least in part by the strong correlation between $M_{\rm\star,Dom}$\ and $M_{\rm\star,Gal}$\ in FIRE (see Figure \ref{fig:sat-residuals}). It is also important to note that the dynamic range of $M_{\rm\star,Dom}/M_{\rm\star,Gal}$\ for the FIRE galaxies is even smaller than $M_{\rm\star,Dom}$: nearly all of the FIRE galaxies exhibit $M_{\rm\star,Dom}/M_{\rm\star,Gal}$\ in between the values for the MW (0.05) and M31 (0.2), compared to the range $M_{\rm\star,Dom}/M_{\rm\star,Gal}$\,=\,0.008--0.5 for the observed sample.

\section{Updated Accreted Masses and Other FIRE Data}
\label{sec:fire-table}

While \cite{sanderson2018} measured accreted masses for the FIRE sample, there were some differences in the mass resolution of the galaxies and the subgrid physics used between different simulated galaxies. The runs from which the satellite catalogs adopted in this paper were created \citep{samuel2020} to include some of the same systems from \cite{sanderson2018} that have since been re-simulated at higher resolution, and all now include a subgrid prescription for turbulent metal diffusion not present in the original simulations. This affects both the isolated `m12' suite, which achieves the now-standard FIRE initial mass resolution of 7070 $M_{\odot}$\ for star and gas particles, and the Local Group-analogous `ELVIS on FIRE' suite, which was resimulated at twice the resolution. See \cite{garrison-kimmel2019} for more nuanced details of the updated simulations. 

With these updates to the models, there is concern that the orbital properties of mergers may be slightly different than previous runs, and therefore differences could be present in the distribution of accreted material relative to the measurements of \cite{sanderson2018}. To address this, we use the star particle catalogs for each FIRE galaxy to estimate updated accreted masses. We use the same scheme as \cite{sanderson2018} to effectively `tag' accreted particles; mainly, that they initially formed at a distance greater than 30\,kpc from the host ($d_{\rm form}\,{>}\,30$\,kpc) and do not belong to an identified satellite at $z\,{=}\,0$. This selection scheme effectively separates accreted and \textit{in situ} material, accurately accounts for the steep observed accreted density profiles and consequent dominant reservoir of accreted material at small radii, and produces accreted masses that are consistent with estimates using the observational techniques described in \S\,\ref{sec:sh-bg}. 

In Table \ref{tab:fire} we present the updated accreted masses, dominant merger masses, and sightline-averaged satellite counts for galaxies in the FIRE simulation, which were used in our comparison to the observations. A description of the calculation of $N_{\rm Sat}$\ can be found in \S\,\ref{sec:model-comparison}. We stress that while future works are fully free to use and distribute the updated accreted stellar masses presented here, \cite{sanderson2018} should also be cited, as we adopt the particle-tagging method devised in that work for identifying accreted star particles. 

\input{firetbl}

\section{Comparing FIRE and Auriga}
\label{sec:fire-auriga}

Currently, two fully hydrodynamic, high-resolution cosmological zoom-in simulations meet the necessary criteria to compare complete satellite populations and merger histories in a sample of MW-mass galaxies: the FIRE \citep{hopkins2018} and Auriga simulations \citep{grand2016}. FIRE achieves higher resolution for its satellite galaxies, which is why it was chosen as the default simulation for model--observation comparison throughout the main body of the paper. However, Auriga also achieves impressively high resolution, and has simulated an even larger sample of MW-like systems. In this section, we compare these two simulations directly to show their very similar behavior in the context of comparing satellite populations and merger histories. 

Each simulation has resolved both the satellite galaxy populations \citep{simpson2018,garrison-kimmel2019} and stellar halo properties \citep{sanderson2018,monachesi2019} of a sample of MW-mass galaxies, with virial masses ranging from approximately $0.8{-}2{\times}10^{12}\,M_{\odot}$. While FIRE resolves dwarf galaxies down to $M_{\star}\,{>}\,10^{5}\,M_{\odot}$, Auriga only resolves satellites to $M_{\star}\,{>}\,4{\times}10^{5}\,M_{\odot}$. When comparing simulations to each other, we consider dwarf galaxies in FIRE to a limit of $M_{\star}\,{>}\,4{\times}10^{5}\,M_{\odot}$\ to put the simulations on equal footing, noting that there may be other differences between them owing to their modest differences in resolution.

\begin{figure}[t]
\leavevmode

\begin{minipage}{\linewidth}
    \centering
    \includegraphics[width={0.75\linewidth}]{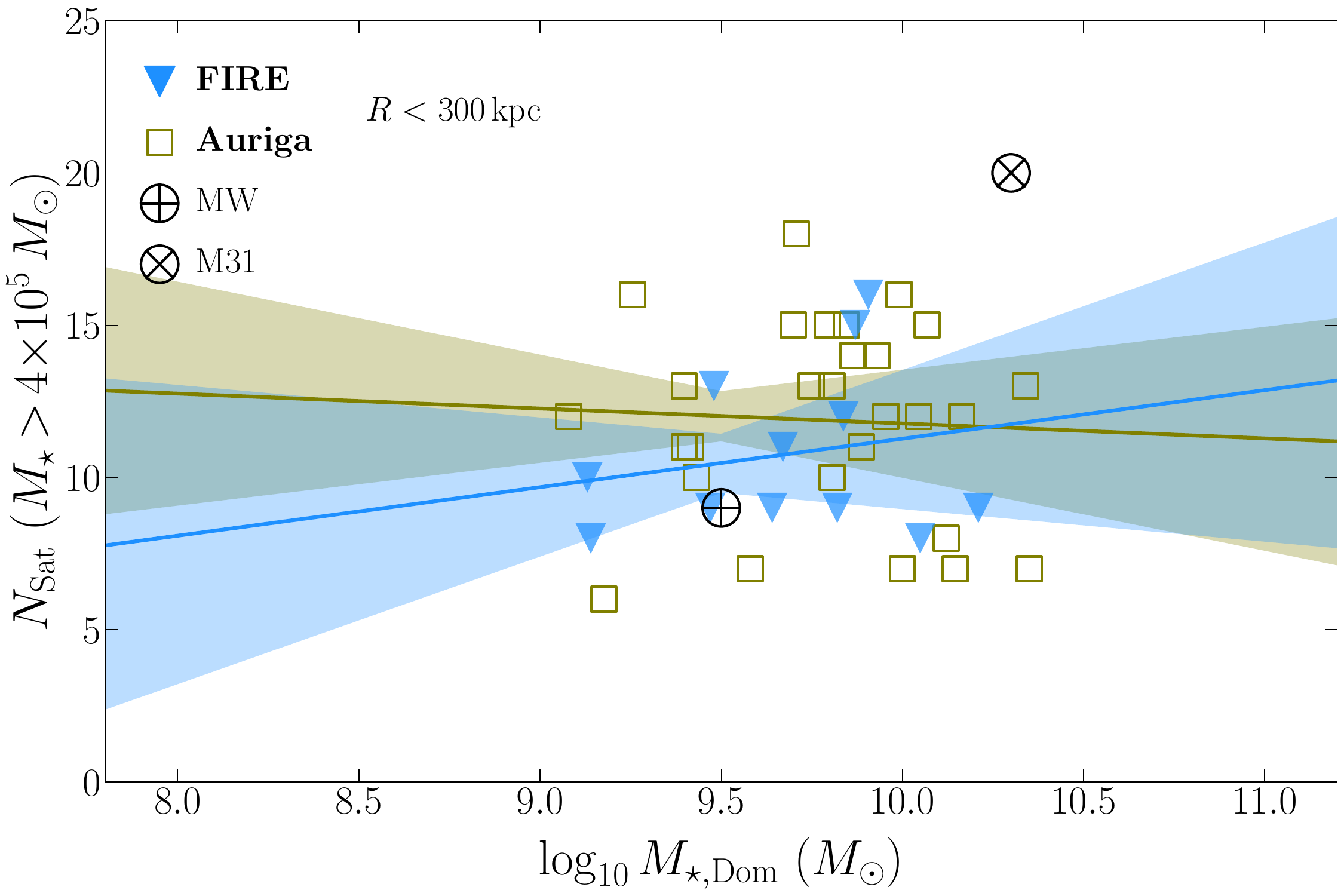}
\end{minipage}

\begin{minipage}{0.48\linewidth}
    \flushright
    \includegraphics[width={0.85\linewidth}]{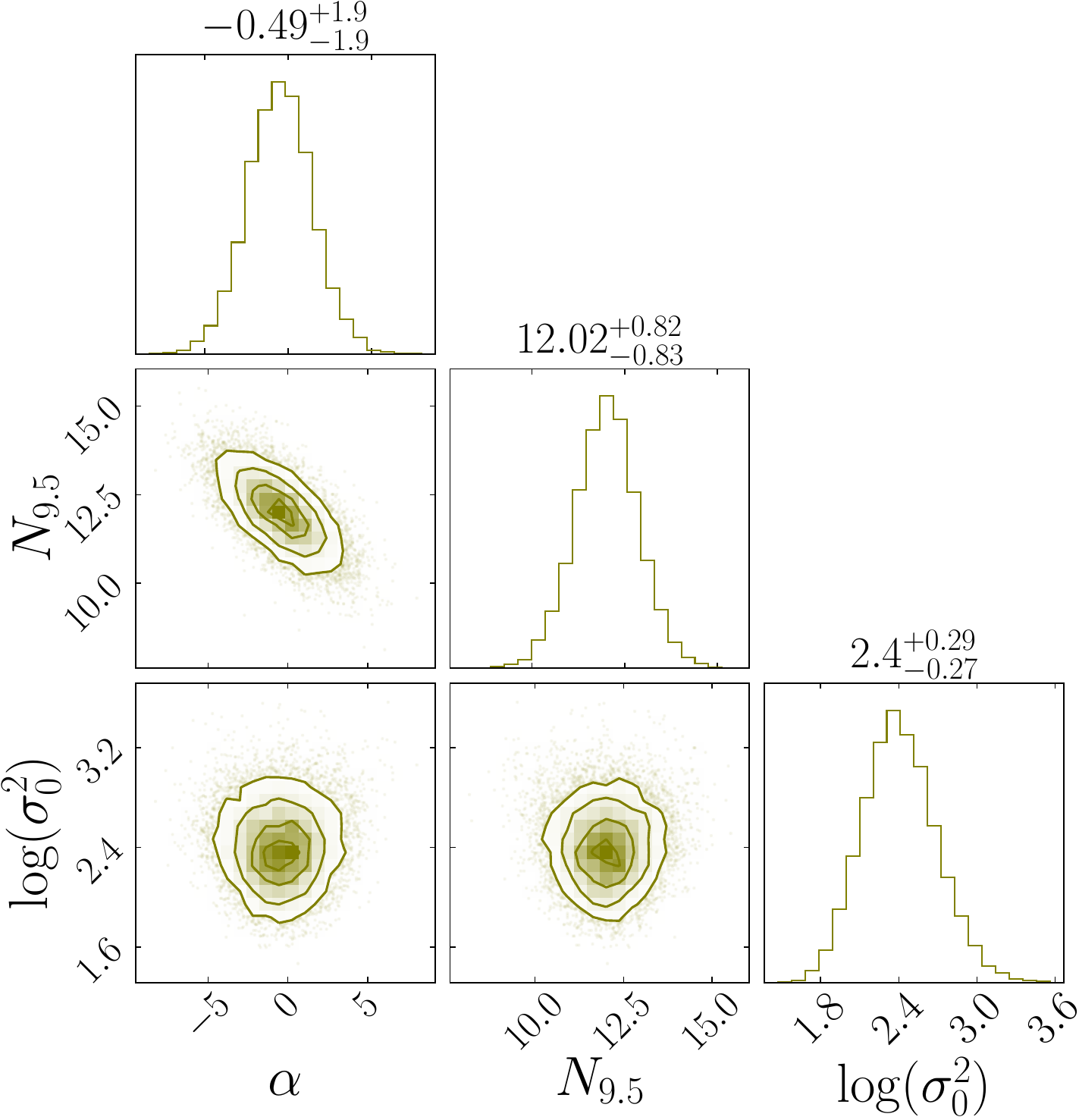}
\end{minipage}
\begin{minipage}{0.48\linewidth}
    \centering
    \includegraphics[width={0.85\linewidth}]{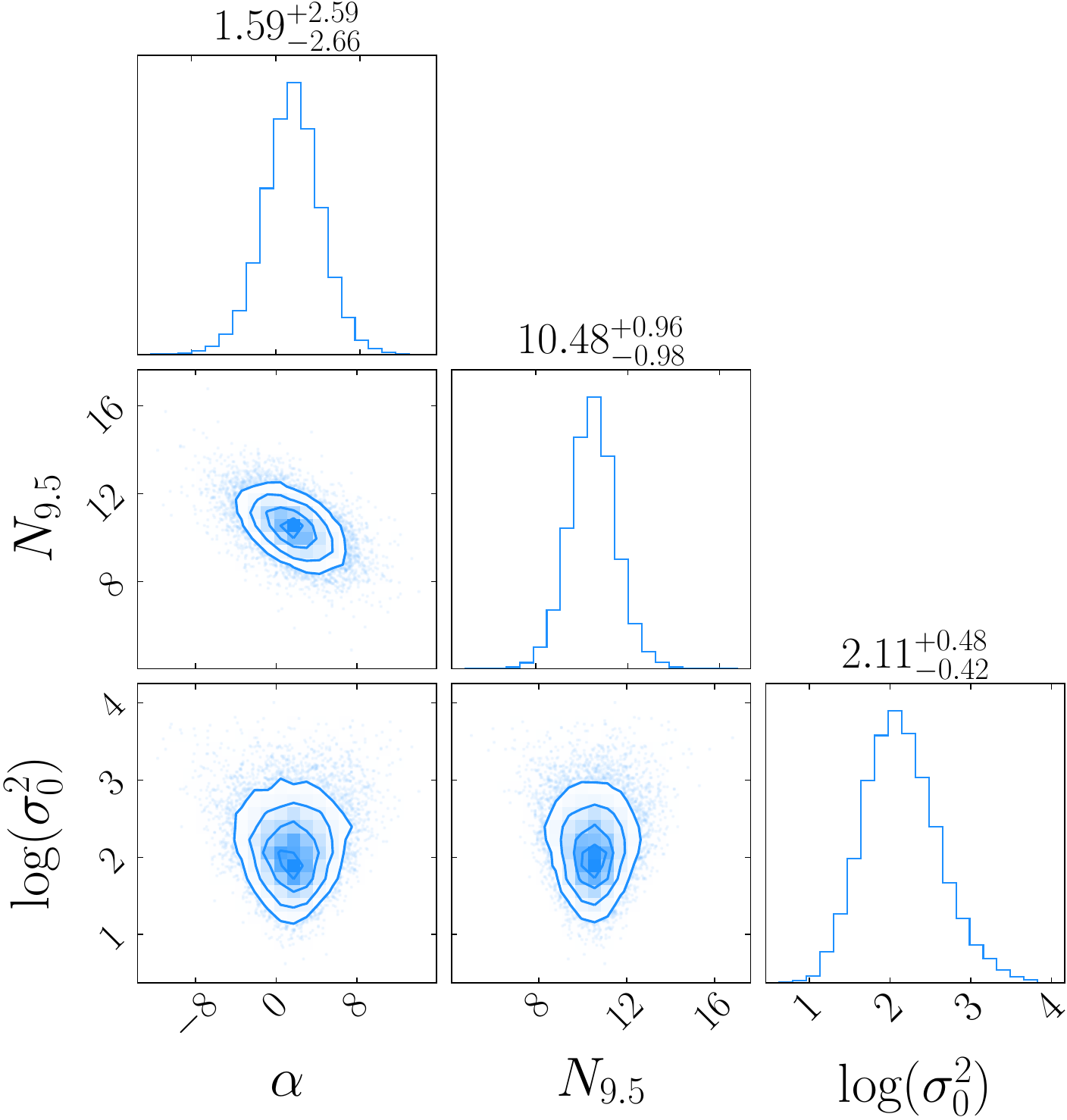}
\end{minipage}
\caption{\uline{Top}: MW-mass systems from the Auriga (green) and FIRE (blue) simulations, showing each systems' estimated $M_{\rm \star,Dom}$\ plotted against the total number of simulated satellites with $M_{\star}\,{>}\,4{\times}10^5\,M_{\odot}$\ within 300\,kpc radius. Data for Auriga were taken from \citet{simpson2018} and \citet{monachesi2019}, while FIRE data were calculated in this paper (see above), using data taken from \citet{sanderson2018}, \citet{garrison-kimmel2019}, and \citet{samuel2020}. We  also show the MW ($\oplus$) and M31 ($\otimes$) for comparison. For each simulation, we show the results of our MCMC analysis, including the models with the likeliest value for each parameter, as well as the 16--84\% confidence envelopes. \uline{Bottom}: Corner plots for FIRE and Auriga from the MCMC analysis. Both simulations are described by very similar model parameters and are consistent with pure scatter --- a near-zero slope and large intrinsic scatter term, in the context of the assumed model.
}
\label{fig:model_sat-accretion}
\end{figure}

In Figure \ref{fig:model_sat-accretion} we show the estimated mass of the most dominant merger (either accreted mass or largest existing satellite) plotted against the number of satellites within 300\,kpc with stellar mass above $4{\times}10^5\,M_{\odot}$, for both the Auriga and FIRE simulations. As in \S\,\ref{sec:model-comparison}, for the FIRE runs we combine the Local Group analog and isolated `m12' suites, as our observational comparison set combines a number of different environments. 

For each simulation suite, we run \texttt{emcee} to evaluate the posterior probability distributions for each of the model parameters defined in Equation \ref{eq:1}, identical to our approach in \S\,\ref{sec:significance} \&\ \ref{sec:model-comparison}. The `best' model, with parameters corresponding to the likeliest value for each marginalized posterior distribution, is shown in Figure \ref{fig:model_sat-accretion} for both FIRE and Auriga, along with their respective 16--84\% confidence regions. Both models are consistent with no intrinsic correlation, with nearly flat slopes and large intrinsic scatter. We show the positions of the MW and M31 for comparison, adopting the $M_{\rm \star,Dom}$\ values given in Table \ref{tab:obs} and $R<300$\,kpc satellite populations compiled by \cite{garrison-kimmel2019}. While M31 is a relative outlier in its satellite populations, compared to both simulations, both FIRE and Auriga closely bracket the MW and M31 in $M_{\rm \star,Dom}$--$N_{\rm Sat}$\ parameter space and possess a very similar range of dominant merger masses. This comparison highlights that different high-resolution models, with different subgrid prescriptions for baryonic physics, consistently meet the benchmark set by satellite populations in the Local Group, and yet fail to capture the emergence of a tight scaling relation between $M_{\rm \star,Dom}$\ and $N_{\rm Sat}$. 

\end{document}

%% file: obstbl.tex
\newcolumntype{t}{!{\extracolsep{7pt}}l!{\extracolsep{0pt}}}

\begin{deluxetable*}{lclccct}[t]
\tablecaption{Nearby Galaxy Group Properties\label{tab:obs}}
\tablecolumns{7}
\tabletypesize{\small}
\tablewidth{250pt}
\tablehead{%
\colhead{Galaxy} &
\colhead{$\log_{10}M_{\rm \star,Gal}$} & 
\colhead{Dom.\,Satellite} &
\colhead{$\log_{10}M_{\rm \star,Dom.\,Sat}$} &
\colhead{$\log_{10}M_{\rm \star,Acc}$} &
\colhead{$N_{\rm Sat}$} &
\colhead{References} \vspace{-5pt} \\
\colhead{} &
\colhead{($M_{\odot}$)} & 
\colhead{} &
\colhead{($M_{\odot}$)} &
\colhead{($M_{\odot}$)} &
\colhead{} &
\colhead{} 
}
\startdata
Milky Way & 10.79 & LMC & \textbf{9.50} & 8.95 & 7\,$\pm$\,1 & 1, 2, 9, 15, 16 \\
Messier 31 & 11.01 & M33 & 9.51 & \textbf{10.3} & 12\,$\pm$\,2 & 2, 10, 16, 25 \\
Messier 81 & 10.75 & M82 & \textbf{10.45} & 9.06 & 11\,$\pm$\,2 & 2, 11, 14, 18, 20, 22 \\
Messier 83 & 10.72 & KK 208 & $\sim$7.81 & \textbf{8.62} & 4\,$\pm$\,1 & 2, 4, 17, 26 \\
Messier 94 & 10.72 & M94-dw1 & 5.99 & \textbf{8.70} & 2\,$\pm$\,0 & 21, 24, 27 \\
Messier 101 & 10.77 & NGC 5474 & \textbf{9.13} & 7.91 & 5\,$_{-1}^{+3}$ & 2, 3, 6, 8, 12, 18, 20 \\
Messier 104 & 11.30 & dw1240-1118 & $\sim$7.66 & \textbf{11.0} & 15\,$_{-4}^{+3}$ & 5, 6, 13, 23 \\
Centaurus A & 11.05 & ESO324-024 & $\sim$8.13 & \textbf{10.6} & 14\,$\pm$\,3 & 2, 7, 10, 19 \\
\enddata
\tablecomments{The central galaxy stellar mass, name and mass of the most massive existing satellite, estimated total accreted stellar mass, and number of satellites within 150\,kpc of the central with $M_V\,{<}\,{-}9$, for each of the 8 nearby MW-mass galaxies composing our observational sample. The larger of the current dominant satellite mass and the accreted mass is bolded, which represents $M_{\rm \star,Dom}$\ for each system. Systems with no published stellar mass for the dominant satellite are denoted by a $\sim$; in these cases we convert the published $M_V$\ by assuming a 1\,$M_{\rm \star}{/}L_V$\ mass-to-light ratio. \textbf{References}: 
(1) \cite{bell2008}, 
(2) \cite{bell2017}, 
(3) \cite{bennet2019}, 
(4) \cite{carrillo2017}, 
(5) \cite{cohen2020}, 
(6) \cite{carlsten2021}, 
(7) \cite{crnojevic2019},
(8) \cite{danieli2017}, 
(9) \cite{deason2019}, 
(10) \cite{dsouzabell2018b}, 
(11) \cite{harmsen2017}, 
(12) \cite{jang2020}, 
(13) \cite{jardel2011},
(14) \cite{karachentsevkudrya2014}, 
(15) \cite{mackerethbovy2020}, 
(16) \cite{mcconnachie2012}, 
(17) \cite{muller2017}, 
(18) \cite{querejeta2015}, 
(19) \cite{rejkuba2011}, 
(20) \cite{sheth2010}, 
(21) \cite{smercina2018}, 
(22) \cite{smercina2020}, 
(23) \cite{tempeltenjes2006},
(24) \cite{trujillo2009},
(25) \cite{vandermarel2012},
(26) Cosby et al., in prep,
(27) Gozman et al., in prep.}
\vspace{-20pt}
\end{deluxetable*}

%% file: m81.tex
\newcolumntype{p}{!{\extracolsep{9pt}}r!{\extracolsep{0pt}}}

\begin{deluxetable}{lccpp}
\tablecaption{M81 Satellites\label{tab:m81}}
\tablecolumns{5}
\tabletypesize{\small}
\tablewidth{250pt}
\tablehead{%
\colhead{Galaxy} &
\colhead{$R_{\rm proj}$} & 
\colhead{$M_V$} & 
\colhead{$\tau_{50}$} & 
\colhead{$\tau_{90}$} \vspace{-5pt} \\
\colhead{} &
\colhead{(kpc)} & 
\colhead{} &
\colhead{(Gyr)} & 
\colhead{(Gyr)} 
}
\startdata
\textbf{BK3N} & 11 & $-$9.59 & 3.9\,$_{-0.5}^{+6.2}$ & 3.2\,$_{-2.7}^{+0.5}$ \\
\textbf{KDG61} & 31 & $-$13.87 & 12.9\,$_{-10.9}^{+0.3}$ & 1.6\,$_{-1.0}^{+2.0}$ \\
\textbf{FM1} & 61 & $-$11.46 & 13.0\,$_{-2.4}^{+0.3}$ & 4.1\,$_{-1.3}^{+8.5}$ \\
BK5N & 72 & $-$11.33 & 13.0\,$_{-5.5}^{+0.0}$ & 10.8\,$_{-9.3}^{+1.7}$ \\
IKN & 82 & $-$11.51 & 11.7\,$_{-3.5}^{+1.1}$ & 10.3\,$_{-4.0}^{+0.8}$ \\
\textbf{KDG64} & 101 & $-$13.43 & 9.3\,$_{-6.3}^{+3.7}$ & 1.8\,$_{-0.5}^{+2.7}$ \\
\textbf{KK77} & 102 & $-$12.84 & 9.8\,$_{-2.0}^{+3.2}$ & 3.4\,$_{-2.0}^{+5.32}$ \\
\textbf{F8D1} & 119 & $-$13.14 & 10.6\,$_{-4.0}^{+4.0}$ & 1.8\,$_{-0.1}^{+10.7}$ \\
\enddata
\tablecomments{50\% and 90\% star formation timescales for faint M81 satellites, calculated from their resolved SFHs \citep{weisz2011}. Satellites are ordered by projected radii, calculated assuming a distance to M81 of 3.6\,Mpc \citep{radburn-smith2011}, and absolute $V$-band magnitudes are given \citep[from][]{karachentsev2000,karachentsev2001,sharina2005,georgiev2009}. Satellites with $\tau_{90}\,{<}\,4$\,Gyr are bolded.}
\vspace{-20pt}
\end{deluxetable}

%% file: firetbl.tex
\newcolumntype{x}{!{\extracolsep{30pt}}c!{\extracolsep{0pt}}}
\newcolumntype{y}{!{\extracolsep{30pt}}l!{\extracolsep{0pt}}}

\begin{deluxetable}{yxxxx}[!h]
\tablecaption{FIRE Data\label{tab:fire}}
\tablecolumns{5}
\tabletypesize{\small}
\tablewidth{250pt}
\tablehead{%
\colhead{Galaxy} &
\colhead{$M_{\rm \star,Gal}$} & 
\colhead{$\log_{10}M_{\rm \star,Dom.\,Sat}$} &
\colhead{$\log_{10}M_{\rm \star,Acc}$} &
\colhead{$N_{\rm Sat}$} \vspace{-5pt} \\
\colhead{} &
\colhead{($10^{10}\ M_{\odot}$)} & 
\colhead{($M_{\odot}$)} &
\colhead{($M_{\odot}$)} 
}
\startdata
m12b & 7.3 & 8.72 & \textbf{10.05} & 6\,$_{-1}^{+2}$ \\
m12c & 5.1 & \textbf{9.91} & 9.59 & 14\,$_{-2}^{+1}$ \\
m12f & 6.9 & 8.22 & \textbf{9.82} & 6\,$\pm$\,1 \\
m12i & 5.5 & 8.10 & \textbf{9.47} & 6\,$\pm$\,1 \\
m12m & 10.0 & 8.94 & \textbf{9.87} & 6\,$_{-1}^{+2}$ \\ 
m12z & 1.8 & 8.81 & \textbf{9.14} & 6\,$\pm$\,1 \\
Romeo & 5.9 & \textbf{9.84} & 9.20 & 10\,$_{-2}^{+1}$ \\
Juliet & 3.4 & 9.30 & \textbf{9.48} & 11\,$_{-3}^{+2}$ \\
Romulus & 8.0 & 8.69 & \textbf{10.21} & 4\,$\pm$\,2 \\
Remus & 4.0 & 8.78 & \textbf{9.64} & 8\,$_{-2}^{+1}$ \\
Thelma & 6.3 & 8.56 & \textbf{9.67} & 9\,$\pm$\,1 \\ 
Louise & 2.3 & 8.56 & \textbf{9.13} & 8\,$_{-2}^{+1}$ \vspace{3pt} \\
\enddata
\tablecomments{Relevant data for the FIRE sample. All data presented here are from the most current highest-resolution runs, and all include the subgrid physics detailed in \cite{garrison-kimmel2019}. \uline{References}: galaxy stellar mass ($M_{\rm \star,Gal}$) --- \cite{samuel2020} \&\ \cite{santistevan2021}; dominant satellite mass ($M_{\rm \star,Dom.\,Sat}$), total accreted stellar mass ($M_{\rm \star,Acc}$), and average number of satellites within projected 150\,kpc ($N_{\rm Sat}$) --- \cite{samuel2020}, \textit{this work}. We ask that any works that wish to use the updated accreted masses also cite \cite{sanderson2018}. 
}
\vspace{-20pt}
\end{deluxetable}